\documentclass[format=acmsmall, review=false, screen=true]{acmart}

\usepackage{booktabs} 

\usepackage{color}

\usepackage{multirow}

\usepackage[ruled]{algorithm2e} 

\SetAlFnt{\small}
\SetAlCapFnt{\small}
\SetAlCapNameFnt{\small}
\SetAlCapHSkip{0pt}
\IncMargin{-\parindent}

\setcopyright{acmcopyright}
\acmJournal{CSUR}
\acmYear{2019}
\acmVolume{1}
\acmNumber{1}
\acmArticle{1}
\acmMonth{1}
\acmPrice{15.00}
\acmDOI{10.1145/3316481}

\begin{document}
\title[Security and Privacy on Blockchain]{Security and Privacy on Blockchain}

\author{Rui Zhang}
\author{Rui Xue}
\affiliation{%
  \institution{State Key Laboratory of Information Security, Institute of Information Engineering, Chinese Academy of Sciences}
  \streetaddress{89 Minzhuang Rd, Haidian Dist}
  \city{Beijing}
  \state{Beijing}
  \postcode{100093}
  \country{China}
}
\affiliation{%
  \institution{School of Cyber Security, University of Chinese Academy of Sciences}
  \streetaddress{19 Yuquan Rd, Shijingshan Dist}
  \city{Beijing}
  \state{Beijing}
  \postcode{100049}
  \country{China}
}
\author{Ling Liu}
\affiliation{%
  \institution{School of Computer Science, Georgia Institute of Technology, USA}
  \city{Atlanta}
  \state{GA}
  \postcode{30332-0765}
  \country{USA}
}

\begin{abstract}
Blockchain offers an innovative approach to storing information, executing transactions, performing functions, and establishing trust in an open environment.
Many consider blockchain as a technology breakthrough for cryptography and cybersecurity, with use cases ranging from globally deployed cryptocurrency systems like Bitcoin, to smart contracts, smart grids over the Internet of Things, and so forth. Although blockchain has received growing interests in both academia and industry in the recent years, the security and privacy of blockchains continue to be at the center of the debate when deploying blockchain in different applications.
This paper presents a comprehensive overview of the security and privacy of blockchain. To facilitate the discussion, we first introduce the notion of blockchains and its utility in the context of Bitcoin like online transactions. Then we describe the basic security properties that are supported as the essential requirements and building blocks for Bitcoin like cryptocurrency systems, followed by presenting the additional security and privacy properties that are desired in many blockchain applications.
Finally, we review the security and privacy techniques for achieving these security properties in blockchain-based systems, including representative consensus algorithms, hash chained storage, mixing protocols, anonymous signatures, non-interactive zero-knowledge proof, and so forth. We conjecture that this survey can help readers to gain an in-depth understanding of the security and privacy of blockchain with respect to concept, attributes, techniques and systems.
\end{abstract}



\maketitle

\renewcommand{\shortauthors}{R. Zhang, R. Xue, L. Liu}

\section{Introduction}
\label{sec:intro}
Blockchain technology is a recent breakthrough of secure computing without centralized authority in an open networked system. From data management perspective, a blockchain is a distributed database, which logs an evolving list of transaction records by organizing them into a hierarchical chain of blocks. From security perspective, the block chain is created and maintained using a peer to peer overlay network and secured through intelligent and decentralized utilization of cryptography with crowd computing.

It is predicted~\cite{coindesk2017Q4} that the annual revenue of blockchain based enterprise applications worldwide will reach \$19.9 billion by 2025, an annual growth rate of 26.2\% from about \$2.5 billion in 2016.
Meanwhile, Goldman Sachs, Morgan Stanley, Citibank, HSBC, Accenture, Microsoft, IBM, Cisco, Tencent, Ali and other world-renowned financial institutions, consulting firms, IT vendors and Internet giants are accelerating laboratory research and capital layout on blockchain technology. Blockchain together with artificial intelligence and big data are considered as the three core computing technologies for the next generation financial industry. In addition to Bitcoin.com, several orthogonal efforts, such as the Hyperledger project sponsored by IBM and Apache foundation, Ethereum~\cite{ethereum,VitalikButerin2014}, FileCoin~\cite{filecoin} provide open source repositories and platforms for blockchain research and development.

Governments have released white papers and technical reports on blockchain to show their positive attitude toward the development of blockchain technology. In UK, the government chief scientific adviser released a new report, which describes the future of distributed ledger technology~\cite{UK:2016:blockchian}. European central bank released documents on distributes ledger technologies in securities post-trading~\cite{ECB:2016}. Chinese government released white paper on blockchain technology and its development in China~\cite{Chinese-whitepaper}. In United States of America (USA), Delaware governor launched ``Delaware Blockchain Initiative", which is a comprehensive program to build a legal and regulatory environment for the blockchain technology development. The state of Delaware governor has officially signed blockchain bill in July, 2017, which, if became law, will formally legitimize and approve those companies registered in the state to manage their accounting and other business transactions using blockchain~\cite{Delaware-blockchain}.

In academia, thousands of papers were published on blockchain in the past five years, including a dozen of study reports on security and privacy threats of blockchain.
Joseph Bonneau et al.~\cite{Bonneau:2015:SoK} provided the first systematic elaboration on Bitcoin and other cryptocurrencies, and analyzed anonymity problems and reviewed privacy enhancing methods.
Ghassan Karame~\cite{Karame:2016:SSB} overviewed and analyzed the security provisioning of blockchain in Bitcoin systematically, including risks and attacks in Bitcoin like digital currency systems. They also described and evaluated mitigation strategies to eliminate some of the risks. Mauro Conti et al.~\cite{Conti:2017:SSP} reviewed security and privacy of Bitcoin, including existing loopholes, which lead to various security risks during the implementation of the Bitcoin system.
Li et al.~\cite{Li:2017:SSB} surveyed the security risks of popular blockchain systems, reviewed the attack cases suffered by blockchain, and analyzed the vulnerabilities exploited in these cases.

Most security and privacy research studies on blockchain have been focused along two threads: (1) uncovering some attacks suffered by blockchain based systems to date, and (2) putting forward specific proposals of employing some state of the art countermeasures against a subset of such attacks. However, very few efforts have been made to provide an in-depth analysis of the security and privacy properties of blockchain and different blockchain implementation techniques. This survey presents a comprehensive review of the security and privacy of blockchains. We first describe the notion of blockchains for online transactions, and discuss the basic and additional security and privacy attributes of blockchains. Then we discuss a set of corresponding security techniques, especially cryptographic solutions, for realizing both basic and additional security goals.
We argue that, as blockchain technology continues to attract attentions and to be deployed in various applications, it is critical to gain an in-depth understanding of the security and privacy properties of blockchain and the degree of trust that blockchain may provide. Such understanding may shed light on the root causes of vulnerabilities in current blockchain deployment models and provide foresight and technological innovation on robust defense techniques and countermeasures.

This survey paper is designed with dual goals. First, it will provide an entry point for non-security experts to gain better understanding of security and privacy properties of blockchain technology. Second, it will help specialists and researchers to explore the cutting edge security and privacy techniques of blockchain. In addition, we identify basic security attributes of blockchain and additional security and privacy properties, discuss some security solutions for achieving these security goals, and insinuate open challenges. We anticipate that this survey will also guide domain scientists and engineers to uncover suitable blockchain models and techniques for many domain specific application scenarios.

We organize the rest of the paper as follows. Section~\ref{sec:overview} describes basic blockchain concepts. Section~\ref{sec:security-properties} describes security attributes that are inherent or desired in blockchain systems. Section~\ref{sec:consensus} introduces consensus algorithms that can be used in blockchain based systems. Section~\ref{sec:privacy-security-tech} discusses the security and privacy techniques that can be employed on blockchain. Section~\ref{sec:conclude} concludes the survey.

\section{Overview of Blockchain}
\label{sec:overview}

The first documented design of blockchain was in 2008, and the first open source implementation of blockchain was deployed in 2009 as an integral element of Bitcoin, the first decentralized digital currency system to distribute bitcoins through the open source release of the Bitcoin peer to peer software. Both were put forward by an anonymous entity, known as Satoshi Nakamoto~\cite{Nakamoto:2008:Bitcoin}.

The Bitcoin system uses the blockchain as its distributed public ledger, which records and verifies all bitcoin transactions on the open Bitcoin peer to peer networked system. A remarkable innovation of the Bitcoin blockchain is its capability to prevent double spending for bitcoin transactions traded in a fully decentralized peer to peer network, with no reliance to any trusted central authority.

{\bf What is Blockchain?\/}
As a secure ledger, the blockchain organizes the growing list of transaction records into a hierarchically expanding chain of blocks~\cite{Narayanan:2016:BCT} with each block guarded by cryptography techniques to enforce strong integrity of its transaction records. New blocks can only be committed into the global block chain upon their successful competition of the decentralized consensus procedure.

Concretely, in addition to information about transaction records, a block also maintains the hash value of the entire block itself, which can be seen as its cryptographic image, plus the hash value of its preceding block, which serves as a cryptographic linkage to the previous block in the blockchain. A decentralized consensus procedure is enforced by the network, which controls (i) the admission of new blocks into the block chain, (ii) the read protocol for secure verification of the block chain, and (iii) the consistency of the data content of transaction records included in each copy of the blockchain maintained on each node.
As a result, the blockchain ensures that once a transaction record is added into a block and the block has been successfully created and committed into the blockchain, the transaction record cannot be altered or compromised retrospectively, the integrity of the data content in each block of the chain is guaranteed, and the blocks, once committed into the blockchain, cannot tampered by any means. Thus, a blockchain serves as a secure and distributed ledger, which archives all transactions between any two parties of an open networked system effectively, persistently, and in a verifiable manner.

In the context of Bitcoin systems, the blockchain is employed as its secure, private and trusted public archive for all transactions that trade bitcoins on the Bitcoin network. This ensures that all bitcoin transactions are recorded, organized and stored in cryptographically secured blocks, which are chained in a verifiable and persistent manner. Blockchain is the pivotal guard in securing bitcoin transactions from many known and hard security, privacy and trust problems, such as double spending, unauthorized disclosure of private transactions, reliance of a trusted central authority, and the untrustworthiness of decentralized computing. The bitcoin way of deploying blockchain has been the inspiration for many other applications, such as healthcare, logistics, education certification, crowd sourcing, secure storage. The blockchain ecosystem is growing rapidly with increasing investment and interests from industry, government and academia.

\subsection{How does the Blockchain Work}
\label{subsec:how}

\begin{figure}
  \includegraphics[width=3.8in]{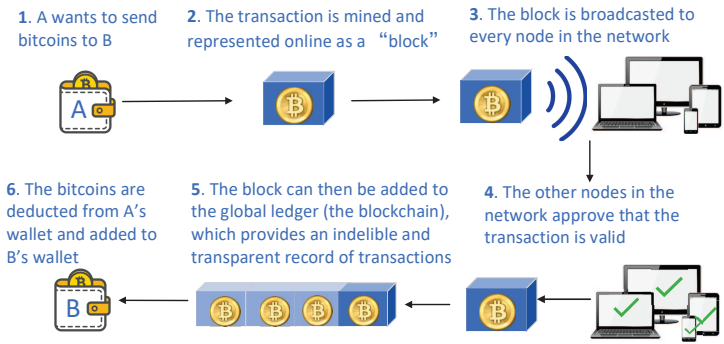}
  \caption{How does the Blockchain Work}
  \label{fig:blockchain}
\vspace{-5mm}
\end{figure}

A blockchain functionally serves as a distributed and secure database of transaction logs. In a Bitcoin network, if client A wants to send some bitcoins to another client B, it will create a bitcoin transaction by client A. The transaction has to be approved by miners before it gets committed by the Bitcoin network. To initiate the mining process, the transaction is broadcasted to every node in the network. Those nodes that are miners will collect transactions into a block, verify transactions in the block, and broadcast the block and its verification using a consensus protocol (a.k.a. Proof of Work) to get approval from the network. When other nodes verify that all transactions contained in the block are valid, the block can be added to the blockchain.
Fig.\ref{fig:blockchain} provides an illustration of this process. Only when the ``block" containing the transaction is approved by the other nodes and added to the blockchain, this bitcoin transfer from A to B will become finalized and legitimate.

Three basic and important capabilities that are supported by the blockchain implementation in Bitcoin are: (1) the hash chained storage, (2) digital signature , and (3) the commitment consensus for adding a new block to the globally chained storage. By an elegant combination of a suite of popular security techniques, such as Hash chain, Merkle tree, digital signature, with consensus mechanisms, the Bitcoin blockchain can prevent both the double spending problem of bitcoins and stop the retrospective modification of any transaction data in a block after the block has been successfully committed into the blockchain.

\subsubsection{Hash Chained Storage}
\label{subsubsec:chained-storage}
Hash pointer and Merkle tree are the two fundamental building blocks for implementing the blockchain in Bitcoin using the hash chained storage.

\paragraph{Hash Pointer}
\label{para:hash-pointer}
Hash pointer is a hash of the data by cryptography, pointing to the location in which the data is stored. Thus, a hash pointer can be used to check whether or not the data has been tampered. A \emph{block chain} is organized using hash pointers to link data blocks together. With the hash pointer pointing to the predecessor block, each block indicates the address where the data of the predecessor block is stored. Moreover, the hash of the stored data can be publicly verified by users to prove that the stored data has not been tampered.

If an adversary attempts to change data in any block in entire chain, in order to disguise the tampering, the adversary has to change the hash pointers of all previous blocks. Ultimately the adversary has to stop tampering because he will not be able to falsify the data on the head of the chain, which is initially generated once the system has been built. We call this initial opening block in the chain the {\em genesis} block. Finally, the adversary's tampering will be uncovered, because by recording this single root hash pointer of {\em genesis} block, one can effectively make the entire chain have the property of tamper-resilient. Users are allowed to go back to some special block and verify it from the beginning of the chain.

\paragraph{Merkle Tree}
\label{para:merkle-tree}
Merkle tree is defined as a binary search tree with its tree nodes linked to one another using hash pointers. It is another useful data structure used for building blockchain. In turn grouping these nodes into disjoint groups, such that each time two nodes at the lower level are grouped into one at the parent level, and for each pair of lower level nodes, the Merkle tree construction algorithm is creating a new data node, which contains the hash value of each. This process is repeated until reaching the root of the tree.

Merkle tree has the ability of preventing data from tampering by traversing down through the hash pointers to any node in the tree. Specifically, when an adversary tries to tamper with data at a leaf node, it will cause a change in the hash value of its parent node, and even if he continues to tamper with the upper node, he needs to change all nodes on the path of the bottom to the top. One can easily detect the data has been tampered with, since the hash pointer of root node does not match with the hash pointer that has been stored.

An advantage of Merkle tree is that it can prove effectively and concisely the membership of a data node by showing this data node and all of its ancestor nodes on its upward pathway to the root node. The membership of Merkle tree can be verified in a logarithmic time by computing hashes on the path and checking the hash value against the root.

\subsubsection{Digital Signature}
\label{subsubsec:signature}

A digital signature establishes the validity of a piece of data by using a cryptographic algorithm. It is also a scheme for verifying that a piece of data has not been tampered with. There are three core components that formulate a digital signature scheme. The first component is the key generation algorithm, which creates two keys, one is used to sign messages and be kept privately and called the private key, and the other is made available to the public, thus called the public key, used to validate whether the message has the signature signed with the corresponding private key. The second core component is the signing algorithm. It produces a signature on the input message endorsed by using the given private key. The third core component is the verification algorithm. It takes a signature, a message, and a public key as inputs, and validates the message's signature using the public key and returns a boolean value.

A well defined and secure signature algorithm should have two properties. The first property is \emph{valid signatures must be verifiable}. The second property is \emph{signatures are existentially unforgeable}. It means that an adversary who has your public key cannot forge signatures on some messages with an overwhelming probability.

\paragraph{Elliptic Curve Digital Signature Algorithm (ECDSA)}
\label{para:ECDSA}
The blockchain used in Bitcoin adopts ECDSA as its digital signature scheme for singing transactions. By employing ECDSA over the standard elliptic curve ``secp256k1", 128 bits of security is provided for Bitcoin blockchain~\cite{Johnson:2001:ECDSA}.
ECDSA proves to be resilient to forgery in the presence of a chosen-message attack based on a generic group and the collision resistant hash function~\cite{Brown:2000:ESE}. Thus, a digital signature scheme like ECDSA should be resistant to a chosen-message attack against a legitimate entity $C$, aiming at fabricating a valid signature on an unseen message $M$, after the adversary obtained the entity $C$'s signature by sending a set of chosen probing queries on a set of messages (not including $M$).

\paragraph{Public Keys as Pseudonyms}
\label{para:pseudonym}
The advantage of using a digital signature is to effectively validate the authenticity of a message by utilizing PKI such that the writer of a message signs it with her private key before sending it out and the recipient of this signed message can use the sender's public key to prove the validity of the message. One can obtain the key pair from a trusted third party in most application scenarios. A PKI is used to manage the public keys via establishing a binding agreement between respective identities of entities (like name, email, and ID) and their public keys. Such binding is done by registering and issuing certificates with a certificate authority (CA). The process of signature verification is automatically translated into identity verification of the signer based on the assurance level of the binding. Thus, public key can be seen as an identity in these scenarios.

While Bitcoin's blockchain adopts decentralizing identity management, without having a central authority to register a user in a system. Key pairs are generated by users themselves. Users can generate key pairs as many as they want. These identities (hashes of public keys) are called addresses in Bitcoin. Because there is no central management of public keys, these identities are actually pseudonyms made up by users.

\subsubsection{Consensus}
\label{subsubsec:consensus}

In the context of decentralized blockchain, when a new block is sent by broadcasting to the network, each node has the option to add that block to their copy of the global ledger or to ignore it. The consensus is employed to seek for the majority of the network to agree upon a single state update in order to secure the expansion of the global ledger (the blockchain) and prevent dishonest attempts or malicious attacks.

Concretely, given that the blockchain is a huge, shared global ledger, anyone may update it, adversarial offense could happen when a node decides to tamper with the state of his copy of the global ledger, or when several nodes collusively attempt on such tampering. For example, if Alice were sending 10 bitcoins to Bob from her wallet, she would like to be sure that no one in the network can tamper the transaction content and change 10 bitcoins to 100 bitcoins.
In order to enable the blockchain to function on a global scale with security and correctness guarantee, the shared public ledger needs an efficient and secure consensus algorithm, which must be fault tolerant, and ensure that (i) all nodes simultaneously maintain an identical chain of blocks and (ii) it does not rely on central authority to keep malicious adversaries from disrupting the coordination process of reaching consensus.
In short, every message transmitted between the nodes has to be approved by a majority of participants of the network through a consensus-based agreement. Also, the network as a whole should be resilient to the partial failures and ``attacks", such as when a group of nodes are malicious or when a message in transit is corrupted.

A good consensus mechanism used in the blockchain implementation also ensures a robust transaction ledger with two important properties: persistence and liveness. Persistence guarantees the consistent response from the system regarding the state of a transaction. For example, if one node on the network states that a transaction is in the ``stable" state, then the other nodes on the network should also report it as stable, if queried and responded honestly. Liveness states that all nodes or processes eventually agree on a decision or a value. By ``eventually", it indicates that it may take a sufficient amount of time for reaching the agreement. By combining persistence and liveness, it ensures that a transaction ledger is robust such that only authentic transactions are approved and become permanent.

In summary, the role of blockchain in the Bitcoin system is to replace the centralized database with authoritative access control. Once some data has been recorded into the global ledger block chain, it should be ``impossible" to change the blockchain, and by enforcing the majority agreement of update validity through consensus, it ensures the consistency state and prevents the double spending problem.
We describe and compare the representative consensus algorithms in Section~\ref{sec:consensus}.

\subsection{Blockchain Level Transaction Models}
\label{subsec:model}
Blockchain is created and maintained as a distributed ledger for online transactions. There are two representative blockchain level transaction models: the unspent transaction outputs (UTXO) model, initially introduced by Bitcoin~\cite{bitcoin} and the account-based transaction model, introduced by Ethereums~\cite{ethereum}. In this subsection, we describe the two transaction models and how their design difference may impact on the solution to the double spending problems.

\subsubsection{The UTXO Model}
\label{subsubsec:utxo}

In Bitcoin and many of its derivatives, a user stores the total amount of her bitcoins as a list of ``unspent" instances of bitcoins that she has received but has not been spent yet. Using the unspent transaction outputs (or UTXO) model, the entire history of the Bitcoin transactions in the system is recorded in a time series of unspent outputs, such that each of them has an owner and a value. The sum of all unspent bitcoin instances that the user has the key to access as the owner in her bitcoin wallet is the total balance of this user. It is straightforward to trace the provenance of individual bitcoins (BTCs) as each of them is signed and sent from one participant to another. A transaction is legitimate if one can prove that the sender has the ownership of the actual bitcoins that are being spent. More specifically, each UTXO transaction can be endorsed if it meets three constraints: (1) Every referenced input in the transaction must be signed by its owner (sender) and not yet spent; (2) If the transaction has multiple inputs, then each input must have a signature matching the owner of the input; (3) A transaction is legal if the total value of its inputs equals or exceeds the total value of its outputs.

Consider an example under the UTXO model: if Bob and Mary both send Alice 5 BTC, and Alice has not spent them, then there are 5 BTC from Bob signed to Alice and another 5 BTC from Mary signed to Alice. If Alice wants to combine her two single instances of 5 BTC into an instance of 10 BTC, Alice must perform another transaction, in a similar way as she would need to exchange her two 5 dollar bills into a 10 dollar bill.

There are a number of benefits for using the UTXO style of online transaction model:
\begin{enumerate}
\item \emph{Potentially high degree of privacy:}
The UTXO model defines a data structure such that each user (the account holder) can hold multiple instances of BTCs without combining them into one total amount, unlike it is done in each of our accounts in a bank. By holding many such instances, account holder need only disclose to her payee (such as Bob) the instances she used to pay the payee. This means that the payer can make multiple payments at the same time. For example, Alice could pay 1 BTC to Bob and 2 BTC to Carol from a 3 BTC instance that Alice holds, without reveal to Bob or Carol the total amount of the aggregate of BTC instances, which Alice has as the owner. Similarly, a user may use different addresses for different transactions that she receives. This will make it difficult to link her accounts to one another.

\item \emph{Potentially high degree of scalability:}
The UTXO model does not have the concept of account for a user, which removes some constrains of the account-based transaction model. A user (payer) can easily send payment transactions in parallel to multiple payees as long as the payer has sufficient granular entries (BTC instances). Such parallel transactions can be executed independently without worrying about transaction ordering, simply because the blockchain relies on hash functions to identify previous states, and thus it is impossible for transactions to be mis-ordered.  As a result, using the UTXO model, one no longer needs to worry about solving the hard problem of keeping track of transaction sequence numbers in a fully distributed system.

\item \emph{Potentially high degree of security:} The UTXO model maintains a Merkle proof of ownership for all BTC instances for each user. Conflict resolution is reduced to the double spending problem, namely, the digital currency-based transactions can easily be duplicated and spent twice. Bitcoin resolves the double spending problem by enforcing a consensus-based confirmation mechanism for committing new blocks into the blockchain and by maintaining the blockchain as a universal ledger.
\end{enumerate}

In a Bitcoin network, the blockchain is created and maintained as a hierarchical and chronologically-ordered chain of blocks with time stamp since its inception in 2009. Each node keeps a copy of the blockchain. A newly created block consisting of several transactions is added to the blockchain. To be secure against double spending, a block should not be considered as confirmed until $\omega$ blocks are added after it (a.k.a. $\omega$ confirmations), The default setting of $\omega$ is six, which means that a transaction contained in the block can be considered as confirmed takes about 60 minutes with the rate of generating a new block is roughly every 10 minutes. In addition, transactions are embedded in blocks and every block is arithmetically linked to the previous block through cryptography. A combination of these techniques makes transactions and blocks immutable and hard to tamper with.

Now we illustrate how the double spending problem is resolved using the Bitcoin blockchain. Assume that Bob sent 1 BTC to Andrew, then signs and sends the same 1 BTC to Alice. Both transactions enter the unconfirmed pool of transactions on the network. If the block containing Bob's first transaction was mined by some miner, and the block containing Bob's second transaction was mined by some other miners. The block containing Bob's first transaction was broadcasted to the entire network and verified by miners first. Then, most miners will continue to mine on the top of the block that containing Bob's first transaction. Thus, Bob's second transaction was judged by the miners as invalid, and pulled from the network.
If both transactions are received by the miners simultaneously, then whichever transaction gets the maximum number of confirmations (blocks deep) first from the network will be included in the blockchain eventually, and the other one will be rejected.

The UTXO model also has some weaknesses, some of which stem from its strengths. For example, if Alice receives 100 BTC and wishes to send Carlo 10 BTC, Alice has to consume the 100 BTC output by creating two outputs: a 10 BTC for the payee Carlo and 90 BTC back to herself as the change. But, the happens of this kind situations may leak private information to an observer. Also, this makes the balance calculation, which is a core feature of the UTXO model, and a significant contributor to wallet's complexity. Although a payer can apply transactions in parallel, it is difficult to achieve them in real parallel due to the need to strictly enforce a total ordering constraint such that the total of the inputs should equal or exceed the total of the outputs.

\subsubsection{Account-Based Online Transaction Model}
\label{subsubsec:account-based}

In contrast to the UTXO model, the account-based online transaction model is by design a simpler model, which explicitly operates all transactions based on the account of senders, instead of unspent transaction outputs, with the objective of improving consensus efficiency and faster block times at the cost of higher degree of risk. It is adopted and extended in Ethereum. Concretely, by the account balance-based transaction model, which operates in a similar way as the bank account in a brick and mortar banking today, a user's entire balance information is stored in Ethereum. A transaction with a token value (ETH) is valid if the following three validity constraints are met: (i) the token is signed by the message writer (sender); (ii) the writer's ownership of token value can be attested, and (iii) the writer's spending account has sufficient balance for the transaction. Upon validation of a transaction, the sending account is debited by the token value and the receiving account is credited with the value.
Thus, a user's account ``balance" in the Ethereum system refers to the sum of the ETH coins for which the user has a private key capable of producing a valid signature. In this model, if Bob has 1 ETH, then upon receiving 1 ETH from Alice, Bob's account balance will be 2 ETH without the need of another exchange transaction to combine the two instances of 1 ETH. In Ethereum, a global state stores a list of accounts with balances, code, and internal storage. It is possible but more difficult to track individual transactions as they are added to the receiver's balance and subtracted from the sender's balance.

There are a number of obvious benefits for the account-based transaction model. First, in contrast to the UTXO model, it has larger space savings, because every transaction in this account balance-based model needs only one reference and one signature to produce an output. Second, it has greater simplicity. Unlike the UTXO model, it does not maintain the source information of coins from transactions in blockchain. Thus, coins are not distinguished based on the sources from which they were received. Third, it does not allow changing reference with each transaction, but it offers easy accessibility to account related data. This is because the Merkle Patricia Tree (MPT) is used to store all account state, transactions and receipts in each block, and a user can scan down the state tree maintained in the MPT along a specific direction to access all data related to an account. In the MPT, SHA3($T$) is used to obtain the hash key of item $T$ in the secure tree (value $T$ being account state, transaction or receipt). As a result, every distinct key/value pair maps uniquely to a root hash, making it very hard to deceive membership of a key/value pair in an MPT.

{\bf Account nonce.\/}
In the account balance-based model, one way to prevent double spends is to have each account associated with a globally accessible nonce, which is simply the count of transactions sending from the account (i.e., the sequence number). Given that this nonce is associated to an account, two accounts may have the same nonce at the same time. Each transaction must assign a ``nonce" to the sending account, which miners check and will process transactions from a specific account in a strict order according to the value of its nonce.
If a block has a transaction with an incorrect nonce, it is considered an invalid block, and other miners will not build on top of it. Hence, if Alice first signs a message sending 100 ETH to Bob, and then signs another message to send 100 ETH to herself from the same account, then using nonce associated to the same sending account of Alice, the second message with higher nonce should not be confirmed before the first. Note that the double spending here is orthogonal to the case in which Alice has two independent accounts, one in Japan and the other in France. Similarly, by associating the transaction counter to a sending account as nonce, the replay attacks can be prevented: namely, a transaction sending 100 ETH from Alice to Bob can be repeated over and over by Bob to continually drain Alice's balance. Thus, maintaining the correct transaction count is very important and failing to increment this value correctly can result in different kinds of errors. For example, reusing nonce or creating incorrect nonce will be detected and rejected: if Alice sends a new transaction for the same account by reusing a past nonce, the mining node will reject the transaction. If Alice sends a new transaction with a nonce that is higher than the correct increment count, the transaction will not be processed until this gap is closed, i.e., until a transaction with each of the missing nonce values has been processed.

{\bf Proof of Work nonce.\/}
In addition to the account nonce, which records the transaction count of an account, Ethereum also uses the proof of work nonce as the second type of nonce, which is the random value in a block that was used to get the proof of work satisfied through mining, an enabling mechanism for decentralized record-keeping.

Ethereum proof of work blockchain is designed in a similar way as that of Bitcoin. A new block can be accepted by the network after being validated through mining. Miners can choose to mine any unverified blocks on the network by solving a puzzle and compete with one another until a winner emerges. If a miner is the first to find a hash that matches the current target, it broadcasts the block across the network to each node. Once the block passes the verification, each node adds this block to their own copy of the ledger. If another miner finds the hash faster, then the rest of miners will stop working on the current block and start the mining process for the next block. This mining process is simultaneously repeated by multiple miners. The consistency is maintained in a decentralized manner by the peer to peer network. The winning miner will be awarded ETH. Similar to Bitcoin, when two miners mine the next block at the same time, the network will decide which one will be the main chain. When two blocks X and Y are mined at the same time. Miners would accept the first block that was broadcasted to them. Thus, some miners accept X and others accept Y. The block that is accepted by the majority of the network (51\% or more) will be the winner. Also, miners who accepted block X will continue to mine the next block on top of block X, and similarly, miners who accepted block Y will continue to mine the next block on top of Y. If the next block is found and added on the top of block X faster, then the miners working on top of block Y will turn to the X chain, which is the main chain. The block X will be the winner and the block Y will become an orphaned block. This decentralized consensus process ensures that any attempt to tamper with the transactions and the blockchain is very hard to fool the majority of the network.

Similar to the PoW in Bitcoin, to counter the domination of the network majority for mining, a system-defined timer is enforced, which controls the hardness of the hash puzzle to ensure that a block can be validated in approximately every 12-15 seconds. If the puzzles are solved faster or slower than this system-default rate, the complexity of the problem is adjusted routinely by the mining algorithm to maintain the roughly 12-second default validation time. This the puzzle-solving method prevents cheating at this game from multiple perspectives, such as leveraging powerful computing resources, forming colluding partners, faking the proof of the correct puzzle answer. Another innovative feature of this mining algorithm is that miners have to find the correct hash value to show the ``proof-of-work", but each node on the network can easily confirm that the hash value is correct. By combining the account nonce with the proof of work nonce, Ethereum can speed up the time required to mine a block significantly compared to Bitcoin, without substantially weakening the resilience of blockchain against malicious manipulation.

\subsection{CAP Properties in Blockchain}
\label{subsec:cap-blockchain}

\subsubsection{CAP Theorem}
\label{subsubsec:cap}
CAP refers to Consistency, Availability and Partition tolerance. CAP theorem is a fundamental theorem for defining transactional properties in distributed systems. A distributed system involves a set of computing nodes that are connected over an overlay network and communicate with one another to accomplish some tasks. CAP theorem states that any distributed systems can have only two of the following properties~\cite{Gilbert:2002:BCF}:
\begin{itemize}
  \item \emph{Consistency:} where each computing node receives the most recent write.
  \item \emph{Availability:} where any requests for some data is always available.
  \item \emph{Partition tolerance:} where the distributed systems is always operational, even when some subset of nodes fail to operate.
\end{itemize}

\subsubsection{CAP Properties in Distributed Ledger - The Problems}
\label{subsubsec:cap-ledger}
In the context of a distributed ledger, CAP properties mean: (1) Consistency: all nodes keep an identical ledger with most recent update. (2) Availability: any transactions generated at any time in the network will be accepted in the ledger. (3) Partition tolerance: even if a part of nodes fail, the network can still operate normally.
The main issue is that it is hard for any widely acceptable currency to exist without all three conditions met. No one will use a currency if the system is not available when the transaction is initiated or some transactions are not recognized by the system. (CP system). No one will use a currency if any one node fails, the system will not operate normally (CA system). No one will use a currency if the ledger saved by different nodes in a distributed ledger system are inconsistent (PA system).

\subsubsection{The Blockchain Solution}
\label{subsubsec:blockchain-solution}
It seems that the CAP theorem has been violated in the blockchain of Bitcoin system, one of the most successful blockchain implementations, because it achieves consistency, availability and partition tolerance. However, this is not the case. In reality, the blockchain consistency is not achieved simultaneously as availability and partition tolerance, but it is after a period of time. The concept of mining is used in Bitcoin, in conjunction with a consensus protocol and a minimum of six confirmations, to ensure eventual consistency through reaching consensus.

\subsection{Classification and Evolution of Blockchains}
\label{subsec:class-evolution}

\begin{table}[t]%
\caption{Classification of Blockchains}
\label{tab:class}
\vspace{-2mm}
\begin{minipage}{\columnwidth}
\begin{center}
\footnotesize
\begin{tabular}{p{1.6cm}p{4.2cm}p{1cm}p{1cm}p{4cm}}
  \toprule
  Types                 &Describe         &\#TA   &SoC            &Scenarios\\ \midrule[0.75pt]
  Public Blockchain     &Anyone can participate and is accessible worldwide&0&Slow&Global decentralized scenarios\\\hline
  Consortium Blockchain &Controlled by pre-selected nodes within the consortium&$\geq 1$&Slight Fast&Businesses among selected organizations\\\hline
  Private Blockchain    &Write rights are controlled by an organization&1&Fast&Information sharing and management in an organization\\
  \bottomrule
\end{tabular}
\end{center}
\bigskip\centering

\end{minipage}
\vspace{-6mm}
\end{table}%

As the blockchain technology continues to evolve with respect to the ways of how blockchains are constructed, accessed and verified, they are being classified into three broad categories: (1) {\em Public blockchain}, which is open for anyone to read, send or receive transactions and allows any participant to join the consensus procedure of making the decision on which blocks contain correct transactions and get added to the blockchain. (2) {\em Consortium blockchain}, which has placed certain constraints on write permissions such that only a pre-selected set of participants in the network can influence and control the consensus process, even though read is open to any participant in the network, and (3) {\em Private blockchain}, whose write permissions are restricted strictly to a single participant (or organization), even though its read permissions are open to the public or constrained to a subset of participants in the network. Although from security and performance perspective, they differ in the speed of consensus (SoC) and whether any of trust authority (TA) is used, and how many of TAs are required, as summarized in Table\ref{tab:class}, these three categories of blockchain share some common properties: (1) they all use decentralized peer to peer networks for transactions; (2) they all require that each transaction is digitally signed and append only to the blockchain, and each peer node maintains a replica of a distributed global ledger of transactions; and (iii) they all rely on consensus to synchronize the replicas across the network.

\begin{figure}
  \includegraphics[width=3.4in]{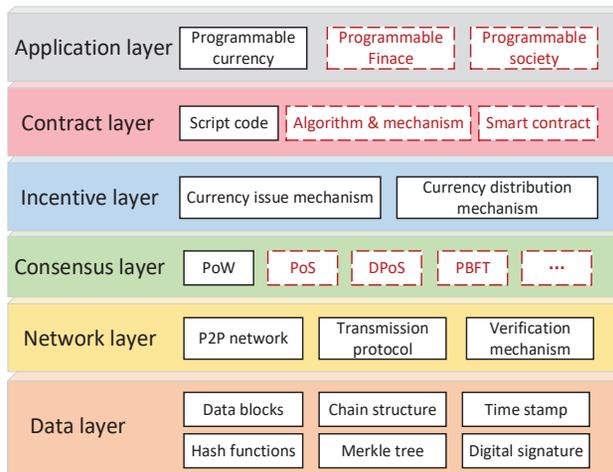}
  \caption{The Architecture of Blockchain}
  \label{fig:architecture}
\vspace{-2mm}
\end{figure}

Although Bitcoin~\cite{Nakamoto:2008:Bitcoin} is publicly released in 2009 as the first peer to peer digital currency system, which implements the blockchain as its public ledger for all transactions, the concept of a secure chain of blocks by cryptography was first proposed in 1991~\cite{Haber1991}, and  the use of Merkle trees as an efficiency optimization of the hash chain was first described by Bayer, Haber and Stornetta~\cite{Bayer1993} in 1993. Over the past 10 years, blockchain has evolved beyond digital currency (Blockchain 1.0), to smart contracts (Blockchain 2.0), and to many other forms of decentralized collaborations with high accountability and high level of security and trust (Blockchain 3.0). Fig.~\ref{fig:architecture} shows the architecture of Blockchain, where the contents of the red dotted were developed by Blockchain 2.0.

As the application utility of blockchain continues to grow from blockchain 1.0 to blockchain 3.0, it becomes even more critical for blockchain users and developers to gain a better understanding of the security and privacy properties of blockchains. We have given an overview of the first implementation of the blockchain in Bitcoin, have touched on some frequently asked questions, such as how the blockchain ensures security through its hash chained storage with hash pointer and Merkle tree and its consensus mechanism, what impact of different online transaction models may have on the security of blockchain, especially against the double spending problem. In the subsequent sections, we focus on discussing the security properties that a blockchain system provides, whether a Bitcoin system can really guarantee anonymity, and what security and privacy techniques can benefit the current and the future generation of blockchain development.

\section{Security and Privacy Properties of Blockchain}
\label{sec:security-properties}
We first discuss the security requirements of online transactions, each of such requirements is targeted at one type of known vulnerabilities.
Then we describe the basic (and inherent) security properties of blockchain based on its first implementation in Bitcoin, and present the set of important additional security and privacy properties of blockchain, which are either present in some existing blockchain systems or desired by many blockchain applications.

\subsection{Security and Privacy Requirements of Online Transactions}
\label{subsec:requirement of online transactions}
We broadly categorize the security and privacy requirements for online transactions into the following seven types.

\subsubsection{Consistency of The Ledger across Institutions}
\label{subsubsec:cross consistency}
In the processes of reconciliation, clearing, and liquidation between financial institutions, due to the architecture and business processes vary from different financial institutions and the involvement of manual processes, it not only leads to high transaction fees generated from the client and the background business side of financial institutions, but also is prone to errors and inconsistencies between ledgers hold by different finance institutes.

\subsubsection{Integrity of Transactions}
\label{subsubsec:integrity}
When using online transactions for investment and asset management, equity, bonds, notes, income vouchers, warehouse receipts, and other assets are managed by different intermediaries. It not only increases the transaction costs, but also brings the risk of deliberately falsifying or forging the certificates. Thus, the system must guarantee integrity of transactions and prevent transactions from being tampered with.

\subsubsection{Availability of System and Data}
\label{subsubsec:avalilability}
The users of online system should be able to access the data of transactions at any time, in anywhere. The availability here refers to both system level and transaction level. At the system level, the system should run reliably even in the event of a network attack. At the transaction level, the data of transactions can be accessed by authorized users without being unattainable, inconsistent, or corrupted.

\subsubsection{Prevention of Double-Spending}
\label{subsubsec:double-spend}
An important challenge in trading digital currency in a decentralized network is how to prevent double-spending, namely spending a coin more than once. In the centralized environment, a trusted central third party is responsible for verifying whether a digital currency has been double-spent or not. For transactions performed in a decentralized network environment, we need robust security mechanisms and countermeasures to prevent double-spending.

\subsubsection{Confidentiality of Transactions}
\label{subsubsec:confidentiality}
In most of the financial online transactions, users wish to have the minimal disclosure of their transactions and account information in an online trading system. The minimal disclosure includes (1) users' transaction information cannot be accessed by any unauthorized user; (2) the system administrator or the participant of the network cannot disclose any user's information to others without his or her permission; (3) all user data should be stored and accessed consistently and securely even under unexpected failures or malicious cyber-attacks. Such confidentiality is desirable in many non-financial scenarios.

\subsubsection{Anonymity of Users' Identity}
\label{subsubsec:id-privacy}
The difficulty of efficient and secure sharing of user data among various financial institutions may result in a high cost of repeated user authentication. It also indirectly brings the disclosure risk of users' identity by some intermediaries.
In addition, one or both parties to the transaction may be reluctant to let the other party know their real identity in some cases.

\subsubsection{Unlinkability of Transactions}
\label{subsubsec:unlinkability}
Different from identity anonymity (not revealing real identity), users should require that the transactions related to themselves cannot be linked. Because once all the transactions relevant to a user can be linked, it is easy to infer other information about the user, such as the account balance, the type and frequency of her transactions. Using such statistical data about transactions and accounts in conjunction with some background knowledge about a user, curious or adversarial parties may guess (infer) the true identity of the user with high confidence.

\subsection{Basic Security Properties}
\label{subsec:basic-defi}
The basic security properties of blockchain stem from both cryptography advances and Bitcoin design and implementation. Theoretically, the first secure chain of blocks was formulated using cryptography in 1991~\cite{Haber1991}. A proposal to improve the efficiency of the cryptographic chain of blocks was put forward in 1993~\cite{Bayer1993}, by incorporating Merkle trees and placing multiple documents into one block. The blockchain is constructed to ensure a number of inherent security attributes, such as consistency, tamper-resistant, resistance to a Distributed Denial-of-Service (DDoS) attack, pseudonymity, and resistance to double-spending attack. However, to use blockchain for secure distributed storage, additional security and privacy properties are required. Table~\ref{tab:relationship} summarizes the set of basic and additional security and privacy properties that need to be ensured for meeting the corresponding requirements outlined in Section~3.1. In the upper part, we show the set of the security and privacy requirements that can be guaranteed by the security and privacy properties and the techniques provided in the original blockchain system, i.e., Bitcoin. In the lower part, we show the security and privacy requirements and properties that need to be strengthened by some additional security and privacy properties and techniques. We describe the basic security and privacy properties in Section~3.2 and the additional properties in Section~3.3. We have briefly mentioned the set of basic security and privacy techniques in Section~2.1, and will detail some of them in Section~4 and will dedicate Section~5 to discuss the additional techniques that can be leveraged to further enhance the security and privacy of blockchains.

\begin{table}%
\caption{Summarization of Security and Privacy Requirements, Properties and Techniques}
\label{tab:relationship}
\begin{minipage}{\columnwidth}
\begin{footnotesize}
\begin{center}
\begin{tabular}{c|p{2.8cm}p{4cm}p{4.5cm}}
  \toprule
\multirow{5}{1.3cm}{\centering Supported in Bitcoin}& S\&P requirements     &S\&P properties & Corresponding S\&P techniques\\\hline
 & Consistency (\ref{subsubsec:cross consistency})     &Consistency (\ref{subsubsec:consistency})&Consensus algorithms (\ref{subsubsec:consensus}, \ref{sec:consensus})\\
 & Integrity (\ref{subsubsec:integrity}) &Tamper-resistance (\ref{subsubsec:tamper-resisitant})&Hash chained storage (\ref{subsubsec:chained-storage})\\
 & Availability (\ref{subsubsec:avalilability})    &Resistance to DDoS attacks (\ref{subsubsec:anti-ddos})&Consensus algorithms with Byzantine fault (\ref{sec:consensus})\\
 & Prevention of double-spending (\ref{subsubsec:double-spend})&Resistance to double-spending attacks (\ref{subsubsec:anti-ds})&Signature and verification (\ref{subsubsec:signature}) \\
& Anonymity (\ref{subsubsec:id-privacy})&Pseudonymity (\ref{subsubsec:pseudonymity})&Public key as pseudonyms (\ref{para:pseudonym})\\\hline
\multirow{3}{1.3cm}{\centering Need to be enhanced}& Unlinkability (\ref{subsubsec:unlinkability})&Unlikability (\ref{subsubsec:unlinkability-property})    &Minxing (\ref{subsec:mixing}), anonymous signature (\ref{subsec:as})\\
&  Confidentiality (\ref{subsubsec:confidentiality}) &Confidentiality (\ref{subsubsec:privacy})  &ABE (\ref{subsec:abe}), HE (\ref{subsec:homomorphic}), SMPC (\ref{subsec:mpc}), NIZK (\ref{subsec:nizk}), TEE-based solutions (\ref{subsec:tee}), game-based solutions (\ref{subsec:game-based})\\
&&Resistance to the majority (51\%) consensus attack (\ref{subsubsec:anti-51})&Consensus algorithms that do not depend on computing power (\ref{sec:consensus})\\
    \bottomrule
\end{tabular}
\end{center}
\end{footnotesize}
\bigskip\centering
\end{minipage}
\vspace{-6mm}
\end{table}%

\noindent
\subsubsection{Consistency}
\label{subsubsec:consistency}
\hspace*{\fill}\\
The concept of consistency in the context of blockchain as a distributed global ledger refers to the property that all nodes have the same ledger at the same time. The consistency property has raised some controversial debate. Some argue that Bitcoin systems only provide eventual consistency~\cite{Wattenhofer:2016:SB}, which is a weak consistency. Other claim that Bitcoin guarantees strong consistency, not eventual consistency~\cite{Emin:2016:BGS}.

Eventual consistency is a consistency model proposed for distributed computing systems by seeking a tradeoff between availability and consistency. Formally, it ensures that all updates to replicas are propagated in a lazy fashion and all read access to a data item will eventually get the last updated value if the item receives no new updates~\cite{Vogels:2009:EC}. In other words, eventual consistency makes sure that data of each entry at each node of the system gets consistent eventually, and thus achieves high availability and low latency at the risk of returning stale data. With eventual consistency, time taken by the nodes of the system to get consistent may not be defined. Thus, data getting consistent eventually means that (1) it will take time for updates to be propagated to other replicas; and (2) if someone reads from a replica which is not updated yet (since replicas are updated eventually), then there is some risk of returning stale data~\cite{Vogels:2009:EC}.

Within a blockchain network system, the strong consistency model means that all nodes have the same ledger at the same time, and during the time when the distributed ledger is being updated with new data, any subsequent read/write requests will have to wait until the commit of this update. In contrast, the eventual consistency model means that the blockchain at each node of the system gets consistent eventually, even though some read/write requests to the blockchain may return stale data. The key challenge for strong consistency is that the performance cost (w.r.t. latency/availability) is too high to be affordable for all cases. The key challenge for eventual consistency is how to remove the inconsistency that may be caused by stale data.
The blockchain in Bitcoin adopts a consistency model that seeks a better tradeoff between strong consistency and eventual consistency for achieving partition tolerance (P) and consistency (C) with deferred availability.

In Bitcoin, transactions are grouped in blocks. When a sender node sends a transaction to the blockchain network, miner nodes will mine it by adding it to a block with other unverified transactions and performing a proof of work challenge game. Upon completing its proof of work challenge, a miner sends its block and its proof to the network to solicit acceptances from other nodes, which will verify all transactions in the block. The other nodes accept the block by working on generating the next block using the hash of the accepted block as its previous hash. The miner whose block is contained in the longest chain and who is the first to obtain $\omega$ confirmations (a.k.a. $\omega$ blocks are appended on the top of the block, and $\omega=6$ by default in Bitcoin consensus protocol) is the winner for chaining this transaction into the distributed global ledger. We can view the $\omega$ parameter as a mechanism to provide configurable or parameterized strong consistency in blockchain.

In summary, blockchain is an elegant approach to addressing the CAP problem for storing a distributed ledger in a decentralized system. For Bitcoin, blockchain implements the partition tolerance (P) while supporting consistency (C) and availability (A) on the clipped blockchain with the most recent $\omega$ blocks disregarded. In short, the consensus protocol accepts an update to the blockchain (the distributed global ledger) only when a number of confirmations received by a miner on its challenge solution is equal to or higher than $\omega$, thus, the update availability is delayed until the $\omega$ confirmations is obtained from the network. The read protocol reads only the blockchain with the last $\omega$ blocks on the chain clipped to ensure the strong consistency and the read availability on the $\omega$-clipped blockchain.
Thus, some has argued that blockchain in Bitcoin guarantees far stronger than eventual consistency. It offers serializability with a probability that is exponentially decreasing with latency~\cite{Emin:2016:BGS}.
On the other hand, certain blockchain applications are less risk-averse and may benefit from a weaker consistency guarantee for convenience and performance. For instance, when $\omega=0$, it means that zero-confirmation is required for both the consensus protocol and the read protocol. This may be a practical choice for those risk-free distributed applications. The blog from Emin G\"un Sirer~\cite{Sirer-blog} is an excellent starting point for more readings on this subject. Furthermore, the time required to confirm a Bitcoin transaction with the $\omega$ constraint for strong consistency may be prohibitively slow for some applications, e.g., 10 minutes on average of generating a block in Bitcoin, and this high latency is aggravated when $\omega$ is configured with higher value. Recently, some research efforts try to build much faster, much higher throughput blockchain systems that provide better guarantees than Bitcoin's 0-confirmation transactions. PeerCensus extends the Bitcoin blockchain to support strong consistency and to decouple block creation and transaction confirmation.

\noindent
\subsubsection{Tamper-Resistance}
\label{subsubsec:tamper-resisitant}
\hspace*{\fill}\\
Tamper-resistance refers to the resistance to any type of intentional tampering to an entity by either the users or the adversaries with access to the entity, be it a system, a product, or other logical/physical object.
Tamper-resistance of blockchain means that any transaction information stored in the blockchain cannot be tampered during and after the process of block generation.
Specifically, in a Bitcoin system, new blocks are generated by mining nodes. There are two possible ways that the transaction information may be tampered with: (1) Miners may attempt to tamper with the information of received transaction; (2) Adversary may attempt to tamper with the information stored on the blockchain. We analyze why such tampering attempts are elegantly prevented by the blockchain protocols in Bitcoin.

For the first kind of tampering, a miner may attempt to change the payee address of the transaction to himself. However, such attempt cannot be succeeded, since each transaction is compressed by a secure Hash function, such as SHA-256, then signed by the payer using a secure signature algorithm, such as ECDSA, in a Bitcoin network, and finally, the transaction is sent to the entire network for verification and approval through mining. Thus, multiple miners may receive and pick up the transaction to mine, which is done in a non-deterministic fashion. If a miner alters any information of the transaction, it will be detected by others when they check the signature with payer's public key, since the miner cannot generate a valid signature on the modified information without the payer's private key. This is guaranteed by the unforgeability of the secure signature algorithm.

For the second kind of tampering, an adversary will fail its attempts to modify any historical data stored on the blockchain. This is because of the two protection techniques used in the distributed storage of blockchain in Bitcoin: the hash pointer, a cryptographic technique, which we mentioned in Section~\ref{subsubsec:chained-storage}, and the network wide support for both storage and verification of the blockchain.
Specifically, if an adversary wants to perform tampering with the data on some block (say $k$), the first difficulty encountered by the adversary is the mismatch problem, namely, the tampered block $k$ has an inconsistent hash value compared to the hash of the preceding block $k$ maintained in the $k+1$ block. This is because using a hash function with collision-resistance, the outputs of the collision-resistent hash function with two different inputs will be completely inconsistent with an overwhelming probability, and such inconsistency can be easily detected by others on the network. Even if the adversary attempts to disguise this tampering by cracking the previous block's hash and so on along the chain, this attempt will eventually fail as the head of the list (a.k.a. genesis block) is reached. Moreover, in the blockchain of Bitcoin network, everyone has a copy of blockchain. It is very hard for an adversary to modify all copies in the entire network.

In short, as every transaction in Bitcoin is signed and distributed over all nodes of the network through the blockchain, it is practically impossible to tamper transaction data without the network knowing about it, showing the power of crowd for storing and distributing the blockchain. This property is attractive to many applications. For example, in healthcare, the blockchain could help to create immutable audit trails, maintain the reliability of health trials, and uphold the integrity of patient data.

\noindent
\subsubsection{Resistance to DDoS Attacks}
\label{subsubsec:anti-ddos}
\hspace*{\fill}\\
A denial-of-service attack refers to as the DoS attack on a host. It is the type of cyber-attacks that disrupt the hosted Internet services by making the host machine or the network resource on the host unavailable to its intended users. DoS attacks attempt to overload the host system or the host network resource by flooding with superfluous requests, consequently stalling the fulfillment of legitimate services~\cite{Mindi:2013:DoS}.

DDoS attack refers to ``distributed" DoS attack, namely, the incoming traffic flooding attack to a victim is originated from many disparate sources distributed across the Internet. A DDoS attacker may compromise and use some individual's computer to attack another computer by taking advantage of security vulnerabilities or weaknesses. By leveraging a set of such compromised computers, a DDoS attacker may send huge amounts of data to a hosting website or send spam to particular email addresses~\cite{Mindi:2013:DoS}. This effectively makes it very hard to prevent the attack by simply jamming individual sources one by one. The arm-race depends on the repairing rate of such compromised nodes against the success rate of compromising computer nodes in the network.

The serious concern in a DDoS attack is on the availability of blockchain and is related to the question of whether a DDoS attacker can make the blockchain unavailable by knocking out a partial or the whole network. The answer to this question is no, thanks to the fully decentralized construction and maintenance of the blockchain and Bitcoin system and the consensus protocol for new block generation and addition to the blockchain, which ensures that the processing of blockchain transactions can continue even if several blockchain nodes go offline. In order for a cyber-attacker to succeed in making blockchain offline, the attacker would have to collect sufficient computational resources that can compromise overwhelmingly large portion of the blockchain nodes across the entire Bitcoin. The larger the Bitcoin network becomes, the harder it is to succeed in such large-scale DDoS attack.

\noindent
\subsubsection{Resistance to Double-Spending Attacks}
\label{subsubsec:anti-ds}
\hspace*{\fill}\\
The double-spending attack in the context of Bitcoin blockchain refers to a specific problem unique to digital currency transactions (recall Section~3.1). Note that the double-spending attack can be considered as a general security concern due to the fact that digital information can be reproduced relatively easily. Specifically, with transactions exchanging digital token, such as electronic currency, there is a risk that the holder could duplicate the digital token and send multiple identical tokens to multiple recipients. If an inconsistency can be incurred due to the transactions of duplicate digital tokens (e.g., double spent the same bitcoin token), then the double-spending problem becomes a serious security threat.

To prevent double-spending, Bitcoin evaluates and verifies the authenticity of each transaction using the transaction logs in its blockchain with a consensus protocol. By ensuring all transactions be included in the blockchain, in where the consensus protocol allows everyone to publicly verify the transactions in a block before committing the block into the global blockchain, ensuring that the sender of each transaction only spends the bitcoins that he possesses legitimately. In addition, every transaction is signed by its sender using a secure digital signature algorithm. It ensures that if someone falsifies the transaction, the verifier can easily detect it. The combination of transactions signed with digital signatures and public verification of transactions with a majority consensus guarantees that Bitcoin blockchain can be resistant to the double-spending attack.

\noindent
\subsubsection{Resistance to the Majority (51\%) Consensus Attack}
\label{subsubsec:anti-51}
\hspace*{\fill}\\
This attack refers to the risks of cheatings in the majority consensus protocol. One of such risks is often referred to as the {\bf 51\% attack}, especially in the context of double-spending. For example, the 51\% attack may occur in the presence of malicious miners. For example, if a miner (verification user) controls more than 50\% of the computing power for maintaining the blockchain, the distributed ledger of all transactions of trading a cryptocurrency. Another example of the 51\% attack may happen when a group of miners collude to perform a conspiracy, e.g., with respect to counting the miners¡¯ votes for verification. If one powerful user or a group of colluding users controls the blockchain, then various security and privacy attacks may be launched, such as illegally transferring bitcoins to some target wallet(s), reversing genuine transactions as if they were never occurred, and so forth.

\noindent
\subsubsection{Pseudonymity}
\label{subsubsec:pseudonymity}
\hspace*{\fill}\\
Pseudonymity refers to a state of disguised identity. In Bitcoin, addresses in blockchain are hashes of public keys of a node (user) in the network. Users can interact with the system by using their public key hash as their pseudo-identity without revealing their real name. Thus, the address that a user uses can be viewed as a pseudo-identity. We can consider the \emph{pseudonymity} of a system as a privacy property to protect user's real name. In addition, users can generate as many key pairs (multiple addresses) as they want, in a similar way as a person can create multiple bank accounts as she wishes.
Although pseudonymity can achieve a weak form of anonymity by means of the public keys, there are still risks of revealing identity information of users. We will further discuss it in Section~\ref{subsubsec:unlinkability-property}.

\subsection{Additional Security and Privacy Properties of Blockchain}
\label{subsec:additional}

Although the blockchain in Bitcoin preserves the three basic security properties: consistency, tamper-resistance, and resistance to DDoS attacks, a general purpose blockchain system may desire and benefit from a set of additional security and privacy properties that are critical to digital currency systems and distributed global ledger services.
Due to space constraint, we here outline a couple of such additional properties.

\noindent
\subsubsection{Unlinkability}
\label{subsubsec:unlinkability-property}
\hspace*{\fill}\\
Unlinkability refers to the inability of stating the relation between two observations or two observed entities of the system with high confidence. Anonymity refers to the state of being anonymous and unidentified. Although the blockchain in Bitcoin ensures pseudonymity by offering pseudo-identity as the support for the anonymity of a user identity, it fails to provide users the protection of unlinkability for their transactions. Intuitively, the full anonymity of a user can only be protected by ensuring both pseudonymity and unlinkability if the user always uses her pseudo-identity to interact with the system. This is because unlinkability makes it hard to launch {\bf de-anonymization inference attacks}, which link the transactions of a user together to uncover the true identity of the user in the presence of background knowledge~\cite{Narayanan:2016:BCT}.
Concretely, in Bitcoin like systems, a user can have multiple pseudonymous addresses. However, this does not provide perfect anonymity for users of blockchain, because every transaction is recorded on the ledger with the addresses of sender and receiver, and is traceable freely by anyone using the associated addresses of its sender and receiver. Thus, anyone can relate a user's transaction to other transactions involving her accounts by a simple statistical analysis of the addresses used in Bitcoin transactions. For example, by analysis on a sender¡¯s account, one can easily learn the number and total amount of bitcoins coming out or going into this account. Alternatively, one can link multiple accounts that send/receive transactions from one IP address. More seriously, a user may lose her anonymity and thus privacy for all the transactions associated with her Bitcoin address if the linkage of her bitcoin address to the user's real-world identity is exposed.

In addition, given the open nature of the public blockchain, anyone can make attempt to perform this type of de-anonymization attack silently and secretly without having the target user even realizing that she is being attacked or her true identity has been compromised. Therefore, the blockchain implementation in Bitcoin only achieves pseudonymity but not unlinkability and thus not full anonymity defined by pseudonymity with unlinkability. We argue that the blockchain system should be enhanced by other cryptographic techniques, outlined in Table~\ref{tab:relationship} and will be discussed in Section~5.

\noindent
\subsubsection{Confidentiality of Transactions and Data Privacy}
\label{subsubsec:privacy}
\hspace*{\fill}\\
Data privacy of blockchain refers to the property that blockchain can provide confidentiality for all data or certain sensitive data stored on it.
Although the blockchain was originally devised as a distributed global ledger for the digital currency system Bitcoin, its potential scope of applications is much broader than virtual currencies. For example, blockchain can be leveraged for managing smart contract, copyrighted works, digitization of commercial or organizational registries.
Not surprisingly, a desirable security property common in all the blockchain applications is the confidentiality of transaction information, such as transaction content (e.g., transaction amounts in Bitcoin), and addresses. Unfortunately, this security property is not supported in Bitcoin systems. In Bitcoin, the transaction content and addresses are publicly viewable, even though the pseudonym is used as the address of sender and receiver of a transaction instead of the real identity. We conjecture that the capability of keeping transaction content private will help to reduce the risk of linkage of pseudonym to the real user identity. This is critical for promoting the need-to-know based sharing instead of publicly viewable of the entire blockchain.

Moreover, blockchain systems, which use smart contracts to implement complex transactions, such as Ethereum, require (1) the data of each contract and the code it runs on the data to be public and (2) every miner to emulate executing each contract. This will lead to the leakage of user information. For example, a user sets up a smart contract to transfer a certain amount of ETH to another user at a certain time. If an adversary has background information about one of the two parties, this adversary may expose and link this party to her real identity. Therefore, it is critical to design and implement stronger protection mechanisms for privacy-preserving smart contracts.

In summary, the data privacy research in the past decades has shown the risks of privacy leakage due to various inference attacks, which link sensitive transaction data and/or pseudonym to the true identity of the real users, even though only pseudonym is being used~\cite{Meiklejohn:2013:FBC,DuPont:2015:TDB}. Such privacy leakage can lead to breaching the confidentiality of transaction information.
Thus, confidentiality and privacy pose a major challenge for blockchain and its applications that involve sensitive transactions and private data. We will dedicate Section~\ref{sec:privacy-security-tech} to discuss some main branches of technology that may help enhancing data privacy and transaction confidentiality on blockchain.

\section{Consensus Algorithms}
\label{sec:consensus}
Consensus is a group-based protocol for reaching agreement dynamically in a group. Compared to the majority voting, a consensus emphasizes that the entire group as a whole could benefit by reaching a consensus.
The problem of dynamically getting a consensus in a group relies on group-based coordination. Such coordinated consensus may be tampered in the presence of malicious actors and faulty processes. For example, a bad actor may secretly create conflicting messages to make group members fail to act in unison, which breaks down the effectiveness of the group to coordinate its actions. This problem is so called the ``Byzantine Generals Problem" (BGP)~\cite{Pease:1980:RAP}. The failure of reaching consensus due to faulty actors is referred to as Byzantine fault. Leslie Lamport, Marshall Pease and Robert Shostak showed in 1982~\cite{Lamport:1982:BGP} that Byzantine fault tolerance can be achieved only if a majority agreement can be reached by the honest generals on their strategy.

The consensus algorithms popularly used in current blockchain systems provide a probabilistic solution to BGP.
In the subsequent sections, we will review these consensus protocols with the focus on their security and privacy properties.\\

\subsection{Proof of Work (PoW)}
\label{para:pow}

The consensus protocol designed by Satoshi Nakamoto~\cite{Nakamoto:2008:Bitcoin} for the Bitcoin is aimed at reaching a coordinated consensus from the network on the validity of each bitcoin transaction. It bypasses the Byzantine Generals Problem by using the PoW protocol.

We characterize the PoW with dual properties: (1) it should be difficult and time-consuming for any prover to produce a proof that meets certain requirement, and (2) it should be easy and fast for others to verify the proof in terms of its correctness. For the first property, one must design a proof of work challenge such that computing a valid proof of work is difficult with low and somewhat random probability, thus a lot of trial and error is needed.

We illustrate how the PoW works in terms of BGP. When the troops on the east of the city want to send a message to the west side troops, it follows the steps of the PoW protocol:
\begin{enumerate}
\item
Append a ``nonce" (usually start with zero) to the original message, which is a random hexadecimal value;

\item
Apply hash to the nonce augmented message and check if the hashing result is less than or equal to a preset value (say starts with five zeros);

\item
If the hash condition is satisfied, the troops on one side of the city will send the messenger to the troops on the other side of the city with the hash of the message and the nonce. If not, then increase the nonce by one and this process iterates until either the desired result is obtained. Finding the right nonce can be time consuming and computationally expensive;

\item Due to the collision-resistant property of hash function, it is hard to tamper the hash of the message even if the messenger got caught, because the hash of the tampered message will be drastically different from the hash of the original message, and the generals on the west of the city can verify whether the message starts with five zeros and disregard the message if it not.

\item
Repeat the above process for multiple iterations such that multiple messengers are sent from the east side troops to the west side troops through the city.
\end{enumerate}

This last step is to address a possible loophole with sending only one messenger: If the city captured the messenger, got the message, tampered with it and then accordingly by changing the nonce until the right nonce value is found such that the desired hash result with required number of zeros is obtained. Even though this process is computationally costly and time consuming, it is still possible. The PoW protocol counters this loophole by increasing strengths in numbers. First, by adding more messengers, the probability of all of them get caught is reduced significantly. Second, even some of them got caught, the amount of time required to tamper the cumulative message and find the corresponding nonce for the hash will be increased substantially. For a block to be valid in the blockchain, a miner has to be able to hash it to a value less than or equal to the current target and then presents its solution to the network for verification by other nodes. The dual properties of PoW ensures that it is extremely difficult and time consuming to find the right nonce for the appropriate hash target; and yet it is super easy and simple to validate the hash result so that no tampering has been made.

The PoW protocol in Bitcoin extends the Hashcash~\cite{Back:2002:Hashcash} system with some minor improvements. First, Bitcoin limits the rate of creating and adding new blocks to the blockchain by the network to roughly one at every 10 minutes. It implements such rate control by automatically monitoring the time spent to solve each proof of work challenge, and adjusting the difficulty of the challenge accordingly. Second, Bitcoin increases the difficulty of predicting which miner in the network will be able to generate the next block by making successful generation of the proof of work at a high cost.

For the formal analysis of PoW, Garay and Kiayias~\cite{Garay:2015:BBP} first formally extracted and analyzed the two fundamental attributes of the Bitcoin protocol: \emph{common prefix} and \emph{chain quality}. However, their analysis is based on several simplifying assumptions, such as a fixed setting with a given number of players, the fully synchronous network channels in which messages are delivered with no delays. Pass et al.~\cite{Pass:2017:ABP} proved that the Nakamoto consensus protocol ensures the blockchain maintaining strong \emph{consistency} and \emph{liveness}, assuming that an asynchronous network has a-priori bounded adversarial delays and the computational challenge is casted as a random oracle.
Recently, Pass and Shi proposed FruitChain~\cite{Pass:2017:Fruitchains}, a protocol that extends the Bitcoin PoW protocol with a reward mechanism, while providing the same consistency and liveness properties with an approximate Nash equilibrium proof.

Although the PoW protocol is effective in solving the Byzantine Generals problem, it suffers three limitations.
{\em First},
the protocol is an extremely inefficient process due to high computation complexity and low probability of successful generation of the proof of work. It is argued that for different applications with different levels of consistency requirements and different risk tolerance levels, it may be attractive to look into more efficient protocols by trade-off between efficiency and strong consistency.
{\em Second},
the proof of work security primarily comes from block creation (mining) rewards, which are strong incentives to attract a large number of miners to participate the proof of work, which is necessary for ensuring the robustness of the PoW blockchain protocol with rigorous security guarantees, defined by \emph{persistence} and \emph{liveness}.
Persistence warrants that as soon as a transaction is appended to a block deep into the blockchain of an honest node such that there are additional $\omega$ or more blocks being placed on top of this block, this transaction will ultimately be contained in every authentic node's blockchain in the network with high probability. Liveness ensures that every transaction originated from an authentic node will be finally stored in a block more than $\omega$ deep of the blockchain of an honest node and become immutable. An honest majority is mandatory for both properties to hold. Although economic consensus has a very important role in protecting liveness and persistence properties in the short term, as shown in the Bitcoin system, seeking moderation by combining with social consensus may hold potential for healthy growth and wide deployment of blockchain in many other applications.
The {\em third} concern is due to the fact that participants may have varying computational capacities and thus different probabilities of successful rates in generating proof of work. It is reported at https://blockchain.info/pools that over 70\% of the hash rate is divided among the top five independent mining farms (e.g., BTC.com 24.4\%, AntPool 14.4\%, ViaBTC 11.1\%, SlusuPool 11.1\%, accessed on March 20, 2018). If these big mining farms were team up with each other, they could acquire more than 51\% of Bitcoin hash power. However, even if an adversary can get unlimited hashing ability, with a 51\% attack of any major blockchain, convincing all nodes of the entire network that this chain is legitimate is much harder than just obtaining the 50\% hash power. Thus, such social layer of consensus may hold some potential towards ultimately protecting any blockchain in the long term.

In summary, the proof of work consensus algorithms tend to rely on anti-centralization incentives and economic incentives for security. By providing block creation rewards to promote more miners, and by requiring to solve computationally expensive challenges to acquire rewards, the former discourages centralized cartels and colluding parties from forming and the latter discourages centralized cartels from acting anti-socially.\\

{\bf Remarks on transaction data tampering attacks.\/}
We would also like to make two final remarks: (1) Even though we have discussed the tampering-proofing techniques in Section 2.1.1 under the hash chained storage, Section 3.2.2 under Tamer-Resistance and Section 4.1 on the Proof of Work. To ease the understanding of the concepts and techniques, in those discussions, we exclude the extreme case of the 51\% attack (i.e., an attacker gains a majority of the hashing power in the network) and dedicate Section 3.2.5 to discuss the 51\% scenario separately.
(2) An adversary may change transaction data in any block (tampering target) on the entire blockchain using at least two tampering methods. Given a tampering target block,  the adversary will need to either change the hashes of all previous blocks (tampering backward) and the hashes of all subsequent blocks of this target block (tampering forward).

\textbf{Backward tampering:} If the adversary chooses to change the previous hashes of blocks, the adversary will fail due to the collision resistant property of hash function and the hash chained structure (even if the adversary can find the second preimage, he needs to continue to change all previous hashes of blocks until reach the genesis block). For this scenario, the security is guaranteed by the hash chained storage and the tamper-resistance property.

\textbf{Forward tampering:} If the adversary chooses to accomplish its attack by performing forward tampering instead, then the adversary needs to redo the proof-of-work for all subsequent (newer) blocks with respect to the target block to obtain a new chain, which needs to be longer than the existing ones, and is chosen as the eventual winner from the network-wide consensus competition. This is because when the hash of a tampering target block is changed, such change will directly impact on the hashes of its subsequent blocks. Thus, by using the forwarding tampering method, the adversary needs to fix all the subsequent blocks of this target tampering block, and also manages to have its tampered chain to be the longest in order to be the winner of the consensus competition for the insertion of this (tampered) chain into the blockchain. Thus, this forward tampering method of transaction tampering attack is related to the proof of work consensus protocol in this section and the 51\% attack (recall Section 3.2.5.). The forward tampering is possible if an adversary has the 51\% hash power for a prolonged period of time.
This forward tampering attack is related to and secured by the proof of work consensus protocol and the assumption that it is very hard if not impossible for an adversary to gain the 51\% hash power. It is also worth to note that the number of blocks to wait for consensus approval before accepting a transaction is captured as a reliably stored parameter ($\omega$), which can be view as a security parameter. The more blocks under which a transaction is stored are, the harder and less feasible it becomes for an adversary to remove it.

\subsection{Proof of State (PoS)}
\label{para:pos}

The proof of stake (PoS) represents an alternative type of distributed consensus protocols for ensuring the CAP properties of public blockchains. It breaks the dependency on rewards for security by promoting penalties-based solutions. Comparing to PoW based public blockchains, such as those used in Bitcoin and the baseline Ethereum, in which any participant on the network can be a miner who validates transactions and creates new blocks to add to the blockchain by solving cryptographical puzzles, PoS based public blockchains, such as the one implemented in Ethereum's Casper, set the constraint on who can be chosen as miners, and by PoS only those participants who have locked up their capital as deposits (stake) are qualified to be chosen as miners or so-called validators in PoS.
The blockchain keeps track of a set of validators that have put aside deposit. Anyone can become a validator by sending a special type of transaction to lock up certain amount of their coins into a validator deposit.
All validators have known identities, which are stable addresses in Ethereum, and the network keeps track of all legitimate validators (those who have reserved coins for participating in validation. Every validator can participate in proposing to create and validate new blocks through a consensus algorithm, which requires the set of validators to put the bet on the next block and take turns to vote and the decision on who will be the validator for the next block is made based on the weight of voting computed by the size of each validator's stake. The probability of being selected is proportional to their bet.

{\bf Rewards v.s. Penalties.\/}
In PoW based blockchain network, miners race to be the first to solve the proof of work, because a reward is given to the first miner who is the winner of adding the next block to the blockchain. This reward includes both the block creation reward and the transaction fees. In contrast, for PoS, to add the next block into the blockchain, each of the qualified validators has to place a bet on the block in order to qualify as a validator for the block. If the block gets appended, then all the validators will get a reward proportional to their bets. There is no block creation award, so validators (miners) are only rewarded by sharing the transaction fees of the block, in addition to the proportion of the bets they put on the block.
Although validators received small rewards in order to compensate them for locking up their state and maintaining nodes and taking additional precaution to secure their private key, the most of the cost of reverting transactions comes from penalties that may thousands of times larger than the rewards that they got in the meantime. Thus, unlike PoW with ``security from rewards of burning computational energy", the proof of stake ensures ``security from penalties of putting up economic value-at-loss".

There are many versions of PoS based consensus algorithms. From an algorithmic perspective, chain-based proof of stake and Byzantine Fault Tolerance (BFT) style proof of stake (see Section~\ref{subsec:BFT}) are the two major types from algorithmic perspective.

In {\bf chain-based PoS}, the algorithm pseudo-randomly selects a validator during each time slot (e.g., every 10 seconds), and assigns that validator the privilege to create a block, and link this block to some previous block (normally the block at the end of the previously longest chain). Thus, over time most blocks converge into a single constantly growing chain.

Early versions of chain-based PoS algorithms are naive because rewards are used for producing blocks with no penalties, and thus suffer from the ``nothing at stake" problem. Concretely, a validator can vote and make blocks on top of multiple competing chains at once and can do so without incurring additional cost. The economically optimal strategy is to vote on as many forks as the validator can find in order to reap more block rewards, because the expected value for voting on multiple competing chains is greater than the expected value for voting on a single chain in such naive PoS design. Even with no attackers, a blockchain may never reach consensus.

In contrast, chain splits are avoided in a PoW system because it is more preferred to add the new block to a longer chain and no miners want to waste resources on a block that will be rejected by the network. Thus, there is an implicit penalty for creating a block on the wrong chain. Also, the ``penalty" for mining on multiple chains is that miners must split up their physical hashing power to do so, or miners have to spend extra electricity and obtain or rent extra hardware. In more recent algorithm proposals for PoS, the strategies for explicit penalties are introduced. For example, one strategy is to penalize those dishonest validators, which create blocks on multiple chains concurrently. If the validators are known and are selected at a time before the fork takes place, then those who place two conflicting signed block headers into the blockchain will be detected as misbehavior. Another strategy is to penalize the validators who create blocks on the wrong chain, and it does not require validators to be known ahead of time.

The Casper PoS protocol in Ethereum is a representative penalty-based PoS protocol. Concretely, the validators put on a portion of their digital currencies (ETHs) as stake for participating in validation of the next blocks. For a new block, only those nodes that want to add it to their local blockchain will place a bet (a portion of their stake) on it and become a validator. The validators can get a reward proportionately to their bets on the block in addition to the transaction fees, only when the block gets verified and appended into the blockchain. If a validator acts maliciously, they will be reprimanded and all of their stakes gets slashed off. The Casper protocol makes malicious players have something to lose and thus the ``nothing at stake" problem is not possible with the Casper consensus protocol.

{\bf Selection of validator for signing the next block.\/}
Although each PoS algorithm defines a way of selecting validators for the next block and signing it to the blockchain, all make effort to avoid undesirable centralization, such as selection by account balance. This is because the single richest member has a permanent advantage of putting the largest deposit of stake. We below describe several different methods for validator selection.

Nxt~\cite{Nxt:2014} and BlackCoin~\cite{BlackCoin:2017} selects the validator as the generator for the next block randomly from those nodes who have put a stake on the block. Randomization based selection adds uncertainty to the decision process that is solely based on the proportion of the stakes. This is because the stakes are public, it is easy to predict which account will likely win the right to generate the next block with reasonable accuracy.

Peercoin~\cite{King:2012:PPCoin} implements its PoS system by combining  randomization with the constraint of ``coin age" to limit the amount of stakes that are used for validation competition. The coin age is defined for each coin by the number of days in which the coin has not been spent. If the constraint of ``coin age" is defined by at least 30 days, then only coins with the coin age of at least 30 days may compete for signing the next block. Those coins that have been used to sign a block must start over with zero ``coin age". One can also set an upper limit of 90 days for the ``coin age". The concept of ``coin age" and the set of constraints make it more difficult to use large stakes to dominate the blockchain. When the size of the network is large, this approach likely makes purchasing more than half of the total stakes in PoS costlier than acquiring 51\% of hashing power in PoW.

Snow White~\cite{Phil:2016:snowwhite} is the first provably secure, robustly reconfigurable consensus protocol for PoS with a growing stakeholder distribution. This protocol proposed a corruption delay mechanism for ensuring security, i.e., robustness under sporadic participation and security in the presence of posterior corruption of past committee members. As long as money does not switch hands too fast (which is enforceable by the cryptocurrency layer), Snow White can attain security when a minority of the stake in the system is controlled by an adversary.

Ouroboros~\cite{Kiayias:2017:Ouroboros} is another PoS blockchain protocol with rigorous security guarantees defined by \emph{persistence} and \emph{liveness}.
By incorporating a reward mechanism into its PoS protocol, Ouroboros proves that honest behavior approximates Nash equilibrium and thus adversarial behavior such as selfish mining can be neutralized and authentic transactions will be approved and become permanent.

\subsection{BFT based Consensus Algorithms}
\label{subsec:BFT}
Byzantine fault tolerance (BFT) is defined as the failure tolerance capability of a system against the Byzantine Generals' Problem (BGP)~\cite{Pease:1980:RAP}. Consider an agreement scenario among a set of players: each player holds a possibly different initial value, and all players need to agree on a single value by obeying a consensus protocol. In a system where such agreement is reached if a majority of the players are honest players who rigorously follow the protocol, even when a minority of the players are malicious and may diverge from the protocol arbitrarily. We consider this system Byzantine fault tolerant.

Most traditional distributed computing systems have central authorities that coordinate and determine what to do next when Byzantine failures occur.
However, in a decentralized blockchain system, there is no central authority. The blockchain is maintained as a distributed global ledger by the network such that each node has a replica of the chain. The initial values are the candidate blocks to be validated and then inserted into the blockchain. For each candidate block, the verification is done by having the network reaching an agreement through the digital signatures of a sufficient number of nodes. Only those candidate blocks that are verified by the network can be added to the blockchain. In order to prevent the occurrence of the Byzantine faults, the blockchain systems rely on the consensus algorithms, such as PoW and PoS, to endorse transactions, which is what makes the blockchain so powerful and so attractive to many applications.

However, neither the PoW nor the PoS algorithm is perfect solution to address the Byzantine fault tolerance (BFT) problem in decentralized peer to peer systems. Understanding BFT is important for applying blockchains solutions to the application areas beyond digital currency. Also, existing consensus algorithms and protocols proposed for the Byzantine fault problem may not be feasible when applied to other blockchain applications. Take healthcare blockchain as an example, asking healthcare providers to spend a lot of computing resources to hashing data is not only very inefficient, but also completely unrealistic. Also, participants in a healthcare blockchain prefer to operate under users with real identities, rather than giving users high anonymity. Thus, relying on node voting and avoiding PoW may be a practical solution to the Byzantine faults.

Since the first solutions to BFT introduced by Lamport, Shostak, and Pease in 1982~\cite{Lamport:1982:BGP}, many Byzantine fault tolerant algorithms and protocols have been put forward that can help resolve many of the misconceptions associated with Byzantine faults and the difficulties in preventing the propagation of related faults.
The PBFT algorithm for practical BFT was proposed by Miguel Castro and Barbara Liskov~\cite{Castro:1999:Practical} in 1999 for high-performance replication of Byzantine state machine. PBFT achieves sub-millisecond increases in latency by processing thousands of requests per second.

AlgoRAND~\cite{Yossi:2017:Algorand} is a new Byzantine agreement protocol that is much more efficient than all previous ones, and has a novel property, called \emph{player replaceability}, which guarantees the security in an adversarial environment. Instead of played by all users in the system, which makes the protocol unwieldy, AlgoRAND chooses its players to be a much smaller subset of users by an algorithm called \emph{cryptographic sortition}, which is a random process of choosing officials from a large set of eligible users. Provided that honest users retain a fraction greater than 2/3 of the money, the main advantages of AlgoRAND are: (1) No forks arise with an overwhelming probability; (2) Only require the minimal amount of computation; and (3) Reach a consensus quickly. Consensus latency is close to the block propagation latency.

HoneyBadgerBFT~\cite{Miller:2016:HBB} is the first practical asynchronous BFT protocol. It is based on a new broadcast protocol that achieves activity and optimal asymptotic efficiency without the need to make any timing assumptions. The main advantage is that HoneyBadgerBFT does not depend on careful tuning of parameters. No matter how unstable the network conditions are, the throughput of HoneyBadgerBFT is always close to the available bandwidth of the network. HoneyBadgerBFT will eventually continue execution once messages are delivered. The authors not only provided the formal proof of the security and liveness of HoneyBadgerBFT protocol, but also showed experimental result that even under optimistic conditions, HoneyBadgerBFT has better throughput than the traditional PBFT protocol~\cite{Castro:1999:Practical}.

\noindent
\subsection{Other Consensus Algorithms}
\label{subsec:others-consensus}

\subsubsection{Sleepy Consensus}
\label{subsubsec:sleepy}
Sleepy consensus~\cite{Rafael:2016:sleepy} considers a ``sleepy model", in which participants are in one of the two modes: ``awake/active" (online) or ``asleep" (offline). Also, participants can change their awake or active status freely during the protocol execution. Sleepy consensus is proved to be resilient when the honest participants are the majority.

Sleepy consensus is established using a Public-Key Infrastructure (PKI).
It proves to be safe and consistent in the context of the Dwork-Naor-Sahai timing model~\cite{Dwork:1998:CZ}, in which the network is assumed to be weakly-synchronous, and there are very small errors between all clocks and the ``real time". Constructed on the collision-resistant hash function, the first protocol of sleepy consensus is easy to implement. But it only supports static corruptions and ``static online schedule". The second protocol of sleepy consensus enhances security in two aspects: (1) it supports adaptive corruptions; (2) it is resilience even under the condition that an adversary arbitrarily chooses which and when the nodes online.
The main idea of Sleepy consensus is established on the Bitcoin PoW blockchain protocol while avoiding using PoW, rather than the standard approaches of distributed consensus. However, Sleepy consensus cannot work in the case of dishonest online players are the majority.

\subsubsection{Proof of Elapsed Time (PoET)}
\label{subsubsec:poet}
Proof of elapsed time (PoET), proposed by Intel~\cite{PoET}, is a blockchain network consensus algorithm that achieves fairness and low computing consumption by leveraging SGX, the Intel trusted computing platform.

In the PoET consensus, each participating node is required to wait for a randomly chosen time period, and the first one to complete the designated waiting time is permitted to generate a new block. Upon broadcasting a new block to the network, SGX facilitates the node to generate an easily verifiable proof of the waiting time.
The PoET consensus needs to ensure two important factors. First, the participating nodes genuinely select a time that is indeed random and not a shorter duration chosen purposely by the participants in order to win. Second, the winner has indeed completed the waiting time.

However, like most of trusted computing technologies, SGX is not entirely reliable. For example, it may not be able to prevent attacks against strategic adversaries equipped with necessary resources. A recent study shows the security vulnerabilities of the blockchain protocol implemented on Intel's SGX platform~\cite{Chen:OSAPoET:2017}. Two countermeasures are proposed to mitigate these vulnerabilities: (1) Altering the probability distribution and (2) Performing statistical tests to reject some blocks that are generated by a given fraction of nodes~\cite{Chen:OSAPoET:2017}.

\subsubsection{Proof of Authority (PoA)}
\label{subsubsec:poa}
Proof of Authority (PoA) is a consensus algorithm that supports comparatively fast transactions~\cite{PoA}. The basic idea of PoA is that only validators have the right to approve transactions and new blocks. A participating node earns reputation to his identity and only when the reputation is accumulated to a high score, the node can become a validator. PoA is considered more robust than PoS for two reasons. On one hand, validators are incentivized to honestly verify transactions and blocks, otherwise their identities will be attached to a negative reputation. On the other hand, a validator cannot approve any two consecutive blocks. This prevents trust from being centralized.

\subsubsection{Proof of Reputation (PoR)}
\label{subsubsec:por}
Proof of Reputation (PoR) can be seen as an extension of PoA. It was recently proposed by different research groups and companies~\cite{Gai:PoR:2018,GoChain:PoR,Aigents:PoR}. The PoR consensus algorithm may have different variations and parameters tuning its performance but the basic idea is simple. Reputation is accumulated and calculated by predefined formulas. Once a node proves reputation and passes verification, it may be voted into the network as an authoritative node and at this point, it operates like a PoA, where only authoritative nodes can sign and validate blocks.

\begin{table}%
\caption{Summary of Consensus Algorithms}
\label{tab:consensus}
\vspace{-2mm}
\begin{minipage}{\columnwidth}
\begin{tiny}
\begin{center}
\begin{tabular}{p{1.2cm}p{1.1cm}p{1.2cm}p{1.5cm}p{1.2cm}p{1cm}p{1.4cm}p{2.6cm}}
  \toprule
  Consensus  &Consistency   &Efficiency  &Resource consumption  &Fault tolerance   &Scalability   &Applications   &Applicable blockchain type\\\hline
  PoW        &Fork          &Low         &Huge                  &$<50\%$             &Poor        &Bitcoin~\cite{Nakamoto:2008:Bitcoin}&\multirow{3}{*}{Public blockchain}\\
  PoS        &Fork          &Higher      &Slight small          &$<50\%$             &Good        &PPcoin\cite{King:2012:PPCoin}&\\
  DPoS       &Fork          &High        &Slight small          &$<50\%$             &Good        &Bitshares\cite{BitShares}&\\\hline
  PBFT       &No fork       &Very high   &Very small            &$<33\%$             &Poor        &Fabric\cite{Fabric}&Consortium blockchain\\\hline
  Paxos/Raft &No fork       &High        &Small                 &$<50\%$            &Good        &Zookeeper\cite{ZooKeeper}&Private blockchain\\
  \bottomrule
\end{tabular}
\end{center}
\end{tiny}
\centering
\end{minipage}
\vspace{-6mm}
\end{table}%

\noindent
\subsection{Comparison of Consensus Algorithms}
\label{subsec:consensus-compare}
In the practical applications, according to different constraints, there are mainly two types of consensus algorithms: strong consistency consensus and eventual consistency consensus.
Typical strong consistency consensus algorithms include BFT and PBFT that consider Byzantine faults as well as Paxos~\cite{Lamport:2016:Paxos} and Raft~\cite{Ongaro:2014:SUC} without considering Byzantine failure.
The strong consistency consensus algorithms are mostly used in the private blockchain and consortium blockchain where the number of nodes is relatively small and there is a stronger requirement for consistency and correctness. More specifically, PBFT is more suitable for consortium blockchain, whereas Paxos and Raft are suitable for private blockchain. Because in both Paxos and Raft, all members of the network are trusted. As shown in Table~\ref{tab:consensus}, Paxos and Raft have better fault tolerance than PBFT, but it should be noted that the fault tolerance of Paxos/Raft only considers node failures without considering the Byzantine fault tolerance. In addition, the communication complexity of Paxos and Raft are much lower than that of PBFT, and they are easy to implement in practical systems.

Typical eventual consistency consensus algorithms include PoW, PoS and DPoS. The eventual consistency consensus algorithms are used in public blockchain where the number of nodes is large and it is difficult to achieve 100\% consistency and correctness for all nodes.
The choice of the consensus algorithm is highly related to the application scenario. As the blockchain technology is applied to various fields, more consensus algorithms suitable for different application fields will be developed in the future.

\section{Privacy and Security Techniques Used in Blockchain}
\label{sec:privacy-security-tech}
In this section, we provide a detailed discussion on a selection of techniques that can be leveraged to enhance the security and privacy of existing and future blockchain systems.

\subsection{Mixing}
\label{subsec:mixing}

As we mentioned before, Bitcoin's blockchain does not guarantee anonymity for users: transactions use pseudonymous addresses and can be verified publicly, thus anyone can relate a user's transaction to her other transactions by a simple analysis of addresses she used in making bitcoin exchanges. More seriously, when the address of transaction is linked to the real world identity of a user, it may cause the leakage of all her transactions. Thus, \emph{mixing services} (\emph{or tumblers}) was designed to prevent users' addresses from being linked. Mixing, literally, it's a random exchange of user's coins with other users' coins, as a result, for the observer, their ownership of coins are obfuscated. However, these mixing services does not provide protection from coin theft. In this section we describe two of such mixing services and analyze their security and privacy properties.

\subsubsection{Mixcoin}
\label{subsubsec:minxcoin}
\hspace*{\fill}\\
Mixcoin~\cite{Bonneau:2014:mixcoin} was proposed by Bonneau et al. in 2014, which provides anonymous payment in Bitcoin and bitcoin-like cryptocurrencies. To defend against passive adversaries, Mixcoin extends the anonymity set to allow all users to mix coins simultaneously. To defend against active adversaries, Mixcoin provides anonymity similar to traditional communication mixes. In addtion, Mixcoin uses an accountability mechanism to detect stealing, and it shows that users will use Mixcoin rationally without stealing bitcoins by aligning incentives~\cite{Bonneau:2014:mixcoin}.

\subsubsection{CoinJoin}
\label{subsubsec:coinjoin}
\hspace*{\fill}\\
CoinJoin~\cite{Maxwell:2013:coinjoin} is proposed in 2013 as an alternative anonymization method for bitcoin transactions. It is motivated by the idea of joint payment. Suppose a user wants to make a payment, she will find another user who also wants to make a payment, and they make a joint payment together in one transaction by negotiation. By the joint payment, it significantly reduces the probability of linking inputs and outputs in one transaction and tracing the exact direction of money movement of a specific user.

CoinJoin requires that users negotiate transactions to whom they wish to joint payment. The first generation of the mixing services to offer this functionality (such as SharedCoin~\cite{sharedcoin}) has used centralized servers and required users to trust the service operator not to steal or allow others to steal the bitcoins. However, despite the single point of failure, centralized services may have risk of leakage of users' privacy, because they will keep logs of the transactions and record all participants of joint payment.

In addition, incorrectly implementation of CoinJoin protocol also will diminish the anonymity. Kristov Atlas identified such flaw in the SharedCoin mixing service~\cite{sharedcoin} and provided a detailed analysis of the flaw in \cite{Atlas:2014:WPG}, In \cite{Atlas:2014:WPG}, Kristov Atlas developed a tool, named ``CoinJoin Sudoku"~\cite{Sudoku}, which could identify SharedCoin transactions and discover relationships between specific payments and payees, indicating that the SharedCoin mixing service can not provide strong privacy for transactions.

CoinShuffle~\cite{Ruffing:2014:CoinShuffle} was proposed by Tim Ruffing et al. in 2014, which further extends the CoinJoin concept and increases privacy by avoiding necessary of trusted third-party for mixing transactions. CoinShuffe is claimed as a completely decentralized coin-mixing protocol and has ability to ensure security against theft. To ensure anonymity, CoinShuffle uses a novel accountable anonymous group communication protocol, which is called Dissent~\cite{Corrigan-Gibbs:2010:Dissent}.

\subsection{Anonymous Signatures}
\label{subsec:as}

Digital signature technology was developed several variants. Some signature schemes themselves has the ability of providing anonymity for the signer. We call this kind of signature schemes anonymous signature. Among the anonymous signature schemes, group signature and ring signature were proposed earlier and are the two most important and typical anonymous signature schemes.

\subsubsection{Group Signature}
\label{subsubsec:gs}

Group signature is a cryptography scheme proposed initially in 1991~\cite{Chaum:1991:GS}. Given a group, any of its members can sign a message for the entire group anonymously by using her personal secret key, and any member with the group's public key can check and validate the generated signature and confirm that the signature of some group member is used to sign the message. The process of signature verification reveals nothing about true identity of the signer except the membership of the group.

Group signature has a group manager who manages adding group members, handling the event of disputes, including revealing the original signer.
In blockchain system, we also need an authority entity to create and revoke the group and dynamically add new members to the group and delete/revoke membership of some participants from the group.

Since the group signature requires a group manager to setup the group, group signature is suitable for consortium blockchain. Recently, JUZIX~\cite{JUZIX} added group signature in its platform for providing users with anonymity support.

\subsubsection{Ring Signature}
\label{subsubsec:rs}

Ring signature~\cite{Rivest:2001:HLS} also can achieve anonymous through signing by any member of a group users. The term of ``ring signature" originates from the signature algorithm that uses the ring-like structure. The ring signature is anonymous if it is difficult to determine which member of the group uses his/her key to sign the message.

Ring signatures differ from group signatures in two principal ways: First, in a ring signature scheme, the real identity of the signer cannot be revealed in the event of dispute, since there is no group manager in ring signature.
Second, any users can group a ``ring" by themselves without additional setup. Thus, ring signature is applicable to public blockchain.

One of typical applications of ring signature is CryptoNote~\cite{CryptoNoteSig:2012}. It adopts ring signature to hide the connection between the sender's addresses of transactions. More precisely, CryptoNote constructs the sender's public key with several other keys, so that it impossible to identify who actually sent (signed) the transaction. Due to the use of ring signature, if the number of ring members is $n$, then the probability that an adversary may successfully guess a real sender of a transaction is $1/n$. Later, Ethereum added ring signature in 2015, which gives the users anonymity like CryptoNote currencies such as Monero~\cite{Monero}.

\subsection{Homomorphic Encryption (HE)}
\label{subsec:homomorphic}

Homomorphic encryption (HE) is a powerful cryptography. It can perform certain types of computations directly on ciphertext, and ensure that the operations performed on the encrypted data, when decrypt the computed results, will generate identical results to those performed by the same operations on the plaintext. There are several partially homomorphic crypto-systems~\cite{Rivest:1978:MOD,ElGamal:1985,Paillier:1999:PCB} as well as fully homomorphic systems~\cite{Gentry:2009:FHE,vanDijk:2010:FHE}.

One can use homomorphic encryption techniques to store data over the blockchain with no significant changes in the blockchain properties. This ensures that the data on the blockchain will be encrypted, addressing the privacy concerns associated with public blockchains. The use of homomorphic encryption technique offers privacy protection, and allows ready access to encrypted data over public blockchain for auditing and other purposes, such as managing employee expenses. Ethereum smart contracts provide homomorphic encryption on data stored in blockchain for greater control and privacy.

\subsection{Attribute-Based Encryption (ABE)}
\label{subsec:abe}

Attribute-based encryption (ABE) is a cryptographic method, in which attributes are the defining and regulating factors for the ciphertext encrypted using the secret key of a user. One can decrypt the encrypted data using the users secrete key if her attributes are agreed with the attributes of the ciphertext. The collusion-resistance is an important security property of ABE.
It ensures that when a malicious user collude with other users, he cannot access other data except the data that he can decrypt with his private key.

The concept of attribute-based encryption was proposed in 2005~\cite{Sahai:2005:FIBE} with single authority. Since then, a number of extensions have been proposed to the baseline ABE, including ABE with multiple authorities to generate users' private keys jointly~\cite{Chase:2007:MABE,Jung:2013:PPC,Lewko:2011:DAE}, ABE schemes that support arbitrary predicates~\cite{Garg:2013:ABE,Gorbunov:2013:AEC}.

Attribute based encryption is very powerful yet few applications to date deploy it due to the lack of understanding of both core concepts and efficient implementation. ABE has not yet been deployed in any form on a blockchain for real-time operation to date. In 2011, a decentralized ABE scheme was proposed~\cite{Lewko:2011:DAE} to employ ABE on a blockchain. For example, on a blockchain, permissions could be represented by ownership of access tokens. All nodes in the network, which have a certain token issued to them, will be granted access to the special rights and privileges associated with the token. The token provides a means of tracking who has certain attributes and such tracking should be done in an algorithmic and consistent fashion by the authority entity that distributes the token. Tokens can be viewed as badges which represent attributes or qualifications, and should be used as non-transferable quantifiers of reputation or attributes.

In \cite{Lewko:2011:DAE}, it is shown that there is no need of a fixed authority to do attribute based encryption. It is possible to have multiple authorities in a decentralized network and fulfill the same accomplishment. For instance, relying on witnesses for the role of these authorities may be possible in a blockchain, with technologies, recently made possible, such as Steemit~\cite{Steemit}, Storj~\cite{storj}, IPFS~\cite{ipfs}, SAFE Network~\cite{SAFE}, though an implementation of attribute based encryption utilizing a blockchain approach remains to be an open challenge.

\subsection{Secure Multi-Party Computation}
\label{subsec:mpc}

The multi-party computation (MPC) model defines a multi-party protocol to allow them to carry out some computation jointly over their private data inputs without violating their input privacy, such that an adversary learns nothing about the input of an authentic party but the output of the joint computation.

Andrew Yao formally defined secure two-party computation in 1982~\cite{Yao:1982:PSC} and generalized it in 1986~\cite{Yao:1986:GES} for the Millionaires' problem. Goldreich et al. proposed a generalization of the two party computation to the multi-party computation in 1987~\cite{Goldreich:1987:HPM}, assuming that all inputs of the computation and zero-knowledge proofs are parts of secret sharing. This generalization has served as the foundation for many subsequent and increasingly efficient MPC protocols. The success of employing MPC in distributed voting, private bidding, private information retrieval has made it a popular solution to many real-world problems. The first large-scale deployment of MPC was in 2008 for an actual auction problem in Denmark~\cite{Bogetoft:2009:SMC}.

In recent years, MPC has been used in blockchain systems to protect users' privacy. Andrychowicz et al. designed and implemented secure multiparty computation protocols on Bitcoin system in 2014~\cite{Andrychowicz:2014:SMC}. They constructed protocols for secure multiparty lotteries without any trusted authority. Their protocols are able to guarantee fairness for the honest users regardless of how dishonest ones behave. If a user violates or interferes with the protocol then she becomes a loser and her bitcoins are transported to the honest users.

A decentralized SMP computation platform, called Enigma, is proposed in 2015 by Zyskind et al.~\cite{Zyskind:2015:Enigma}. By using an advanced version of SMP computation, Enigma employs a verifiable secret-sharing scheme to guarantee privacy of its computational model. Also, Enigma encodes shared secret data using a modified distributed hash table for efficient storage. Moreover, it leverages an external blockchain as a corruption-resistant recording of events and the regulator of the peer to peer network for identity management and access control. Similar to Bitcoin system, Enigma provides autonomous control and protection of personal data while eliminating the necessity and dependency of a trusted third party.

\subsection{Non-Interactive Zero-Knowledge (NIZK) Proof}
\label{subsec:nizk}

Another cryptographic technology that has powerful privacy-preserving properties is zero-knowledge proofs, proposed in the early 1980s~\cite{Goldwasser:1985:KCI}. The basic idea is that a formal proof can be formulated to verify that a program executed with some input privately known by the user can produce some publicly open output with no disclosure of any other information. In other words, a certifier can prove to a verifier that some assertion is accurate without providing any useful information to the verifier.

As a variant of zero-knowledge proofs, it is shown in~\cite{Blum:1988:NZA} that, with the non-interactive variant of zero-knowledge proofs, coined as NIZK, one can achieve computational zero-knowledge without requiring certifier and verifier to interact at all, provided that the certifier and the verifier shares a common reference string.
In a blockchain application, all account balances are encrypted and stored in the chain. When a user transfers money to another user, he can easily prove that he has sufficient balance for the transfer with zero-knowledge proofs, while without revealing the account balance.

Another variation is the zero-knowledge Succinct Non-interactive ARgument of Knowledge (zk-SNARK) proof, introduced in 2012 by Bitansky and his coauthors~\cite{Bitansky:2012:ECR} and is served as the backbone of the Zcash protocol~\cite{Sasson:2014:ZDA}.
Zcash~\cite{Sasson:2014:ZDA} uses zk-SNARKs~\cite{Groth:2010:SPB,Bitansky:2013:SNA} to verify transactions while protecting users' privacy.

Recently, the Zcash group enhanced the Ethereum contract language to efficiently provide zk-SNARK proofs verification. More specifically, they adopted a snark-verify precompile (like an opcode) to a fork of ``Parity" which uses lib-snark to verify generic proofs.
They also used the new zk-SNARK verifier to enforce an original coin mixing contract, which adopts a simplified version of Zerocash, an academic protocol whose implementation is used to build Zcash. Thus, it is called ``baby" ZoE, stand for Zerocash over Ethereum. The contract allows a user to store discrete amounts (units of ETH) by adding a ``serial number" as a commitment into a Merkle tree, which is maintained by the contract.

\subsection{The Trusted Execution Environment (TEE) Based Smart Contracts}
\label{subsec:tee}
An execution environment is called TEE if it provides a completely isolated environment for application execution, which effectively prevents other software applications and operating system(s) from tampering with and learning the state of the application running in it. The Intel Software Guard eXtensions (SGX) is a representative technology to implement a TEE. For example, Ekiden~\cite{Raymond:2018:Ekiden} is a SGX based solution for confidentiality-preserving smart contracts. Ekiden separates computation from consensus. It performs smart contract computation in TEEs on compute nodes off chain, then uses a remote attestation protocol to validate the execution correctness of compute nodes on chain. The consensus nodes are used for maintaining the blockchain and do not require to use trusted hardware. Enigma~\cite{Zyskind:2015:Enigma} utilizes TEE in its current version to allow users to create privacy-preserving smart contracts using a decentralized credit scoring algorithm. Multiple factors are weighted for credit scoring, such as the number and types of accounts, payment history, credit utilization.

\subsection{Game-Based Smart Contracts}
\label{subsec:game-based}
The game-based solutions for smart contracts verification are very recent developments, represented by TrueBit~\cite{Jason:2017:TrueBit} and Arbitrum~\cite{Harry:2018:Arbitrum}.

TrueBit~\cite{Jason:2017:TrueBit} uses an interactive ``verification game" to decide whether a computational task was correctly performed or not. TrueBit offers rewards to encourage players to check computation tasks and find bugs, such that a smart contract can securely perform a computation task with verifiable properties. In addition, in each round of ``verification game", the verifier recursively checks a smaller and smaller subset of the computation, which allows TrueBit to greatly reduce the computational burden on its nodes.

Arbitrum~\cite{Harry:2018:Arbitrum} has designed an incentive mechanism for parties to agree off-chain on the behavior of virtual machines, so that it only requires the verifiers to verify digital signatures of the contracts. For dishonest parties who try to lie about the behavior of virtual machines, Arbitrum has designed an efficient challenge-based protocol to identify and penalize the dishonest parties. The incentive mechanism of off-chain verification of virtual machine's behavior has significantly improved the scalability and the privacy of smart contracts.

\subsection{Discussion}
\label{subsec:discussion}

\begin{table}%
\caption{Summary of Security and Privacy Techniques}
\label{tab:summary-sp-tech}
\vspace{-2mm}
\begin{minipage}{\columnwidth}
\begin{footnotesize}
\begin{center}
\begin{tabular}{p{1.3cm}p{2cm}p{4.5cm}p{4.5cm}}
  \toprule
  Techniques  &Applications  &Advantages  &Disadvantages\\\hline
  Mixing        &Mixcoin~\cite{Bonneau:2014:mixcoin}, CoinJoin~\cite{Maxwell:2013:coinjoin}      &It can prevent users' addresses from being linked. & The centralized
services may have risk of leakage of users' privacy.\\\hline
  Group signature        &JUZIX~\cite{JUZIX}          &The identity of signer can be hidden among a group of users. In the event of a dispute, the identity of the signer can be revealed.      &Need a trusted third party to act as a manager.\\\hline
  Ring signature        &CryptoNote~\cite{CryptoNoteSig:2012}, Monero~\cite{Monero}, Ethereum~\cite{ethereum}          &The identity of signer can be hidden among a group of users. No need for the participation of any trusted third party.      &In the event of a dispute, the identity of the signer cannot be revealed.\\\hline
  ABE       &None         &It can simultaneously achieve data confidentiality and fine-grained access control.    &The issuance and revocation of attribute certificate in a distributed environment still need to be resolved.\\\hline
  HE       &Ethereum~\cite{ethereum}       &It can achieve privacy-preserving computation by performing computations directly on ciphertext.  &Only some types of operations, such as addition and multiplication, can be efficiently implemented. The computational efficiency of complex functions is very low.\\\hline
  SMPC &Enigma~\cite{Zyskind:2015:Enigma}       &It allows multi-party to carry out some computation jointly over their private data inputs without violating their input privacy.       &Only some simple functions can be supported, and complex functions are less efficient.\\\hline
  NIZK  &Zcash~\cite{Sasson:2014:ZDA}             &User can easily prove that he has sufficient balance for the transfer with NIZK, while without revealing the account balance.            &Less efficient      \\\hline
  TEE-based solutions&Ekiden~\cite{Raymond:2018:Ekiden}, Enigma~\cite{Zyskind:2015:Enigma}   &It can protect the privacy of smart contracts by running them in TEE.   &The compute nodes need to be equipped with a CPU, which has TEE, such as Intel SGX. The attacks on SGX still need to be resolved.  \\\hline
  Game-Based solutions&TrueBit~\cite{Jason:2017:TrueBit}, Arbitrum~\cite{Harry:2018:Arbitrum}   &It encourages partis to verify the correctness of smart contracts through incentives mechanism.  &There is still a risk of being deceived by a malicious user.\\
  \bottomrule
\end{tabular}
\end{center}
\end{footnotesize}
\centering
\end{minipage}
\vspace{-6mm}
\end{table}%

We summarize the pros and cons of each security and privacy technique in Table~\ref{tab:summary-sp-tech}. To achieve security and privacy in a complex blockchain system that needs to meet multiple security and privacy requirements with desired properties, we would like to make the following three remarks: (1) No single technology is a panacea for security and privacy of Blockchain. Therefore, the appropriate security and privacy techniques should be chosen based on the security and privacy requirements and the context of application. In general, the combination of multiple technologies works more effectively than using a single technology. For example, Enigma~\cite{Zyskind:2015:Enigma} combines cutting-edge cryptographic technique SMPC and hardware privacy technology TEE with blockchains to provide computation over encrypted data at scale. (2) There is no technology that has no defects or is perfect in all aspects. When we add a new technology to a complex system, it always causes other problems or new form(s) of attacks. This requires careful attentions on the pitfalls and potential harms induced from integrating some security and privacy techniques into the blockchains. (3) There is always a trade-off between security-privacy and efficiency. We should advocate those techniques that improve the security and privacy of blockchain and at the same time promote the practical deployment of the blockchain applications with acceptable performance.

\section{Concluding Remarks}
\label{sec:conclude}

We have presented a survey on security and privacy of blockchain with a number of contributions. First, we characterized the security and privacy attributes of blockchain into two broad categories: inherent attributes and additional attributes in the context of online transactions. Second, we described the security and privacy techniques for achieving these security and privacy attributes in blockchain-based systems and applications, including representative consensus algorithms, mixing, anonymous signatures, encryption, secure multiparty computation, non-interactive zero-knowledge proof, and secure verification of smart contracts. With growing interest of blockchain in both academic research and industry, the security and privacy of blockchains have attracted huge interests, even though only a small part of the blockchain platforms can achieve the set of abovementioned security goals in practice. We argue that an in-depth understanding of the security and privacy properties of blockchain plays a critical role in enhancing the degree of trust that blockchain may provide and in developing technological innovation on robust defense techniques and countermeasures. We conjecture that developing light-weight cryptographic algorithms as well as other practical security and privacy methods will be a key enabling technology in the future development of blockchain and its applications.

\begin{acks}

The first two authors acknowledge the partial support from National Key R\&D Program of China under Grant No.2017YFB1400700 and National Natural Science Foundation of China under Grant No.:~\grantnum{GS501100001809}{61472414, 61772514, 61602061}. The last author acknowledges the partial support by the National Science Foundation under Grants 1564097 and 1547102, and an IBM faculty award.

\end{acks}

\bibliographystyle{ACM-Reference-Format}
\bibliography{blockchain-bibliography}

\end{document}